\documentclass{appolb}
\usepackage{graphicx}

\newcommand{\bea}{\begin{eqnarray}}
\newcommand{\eea}{\end{eqnarray}}

\usepackage{amsmath,amssymb,graphics}
\usepackage{cite}

\def\Mp{{M_{\rm P}}}
\def\mG{{m_{3/2}}}
\begin{document}
\eqsec  % uncomment this line to get equations numbered by (sec.num)
\title{\bf\Large Baryogenesis, Dark Matter and the \\ 
Maximal Temperature of the Early Universe\thanks{Lectures presented at
the LII Cracow School of Theoretical Physics, 
{\it Astroparticle Physics in the LHC Era}, May 19-27, 2012, Zakopane,
Poland}\\ \mbox{}
%
% you can u\\' to break lines
}
\author{Wilfried Buchm\"uller
\address{Deutsches Elektronen-Synchrotron DESY, Hamburg, Germany}
}
\maketitle
\begin{abstract}
\noindent
Mechanisms for the generation of the matter-antimatter asymmetry and
dark matter strongly depend on the reheating temperature $T_R$, the
maximal temperature reached in the early universe.  Forthcoming
results from the LHC, low energy experiments, astrophysical
observations and the Planck satellite will significantly constrain
baryogenesis and the nature of dark matter, and thereby provide
valuable information about the very early hot universe. At
present, a wide range of reheating temperatures is still consistent
with observations. We illustrate possible origins of matter and dark
matter with four examples: moduli decay, electroweak
baryogenesis, leptogenesis in the $\nu$MSM and thermal leptogenesis. Finally, we
discuss the connection between baryogenesis, dark matter and
inflation in the context of supersymmetric spontaneous $B$-$L$ breaking.
\end{abstract}
  
\section{Introduction}

Let us begin by recalling some temperatures, possibly realized in the
hot early universe, the related cosmic times,
and the connection with microscopic physics at the corresponding
energies:
\begin{itemize}
\item
 $T_R \sim 0.1~{\rm eV}$ [$t \sim 10^{13}~{\rm s}$] \\
Light nuclei and electrons form neutral atoms and the universe becomes
transparent to photons. They decouple from the plasma and are
observable today as cosmic microwave background (CMB).
\item
 $T_R \simeq 0.1\ldots 10~{\rm MeV}$\footnote{In
   natural units $\hbar=c=k_B=1$ one has $1~\mathrm{eV} =
   1.16~10^4\mathrm{K}$.} [$t\simeq 10^{2}\ldots 10^{-2}~{\rm s}$]\\
Light nuclei are formed from protons and neutrons (primordial
nucleosynthesis,  BBN) and neutrinos decouple from the plasma.
\item
$T_R\sim 10~{\rm GeV}$ [$t\sim 10^{-8}~{\rm s}$] \\
Weakly interacting massive particles (WIMPs), the most popular
dark matter candidates, decouple from the plasma. 
\begin{figure}[t]
\centerline{
\includegraphics[width=12cm]{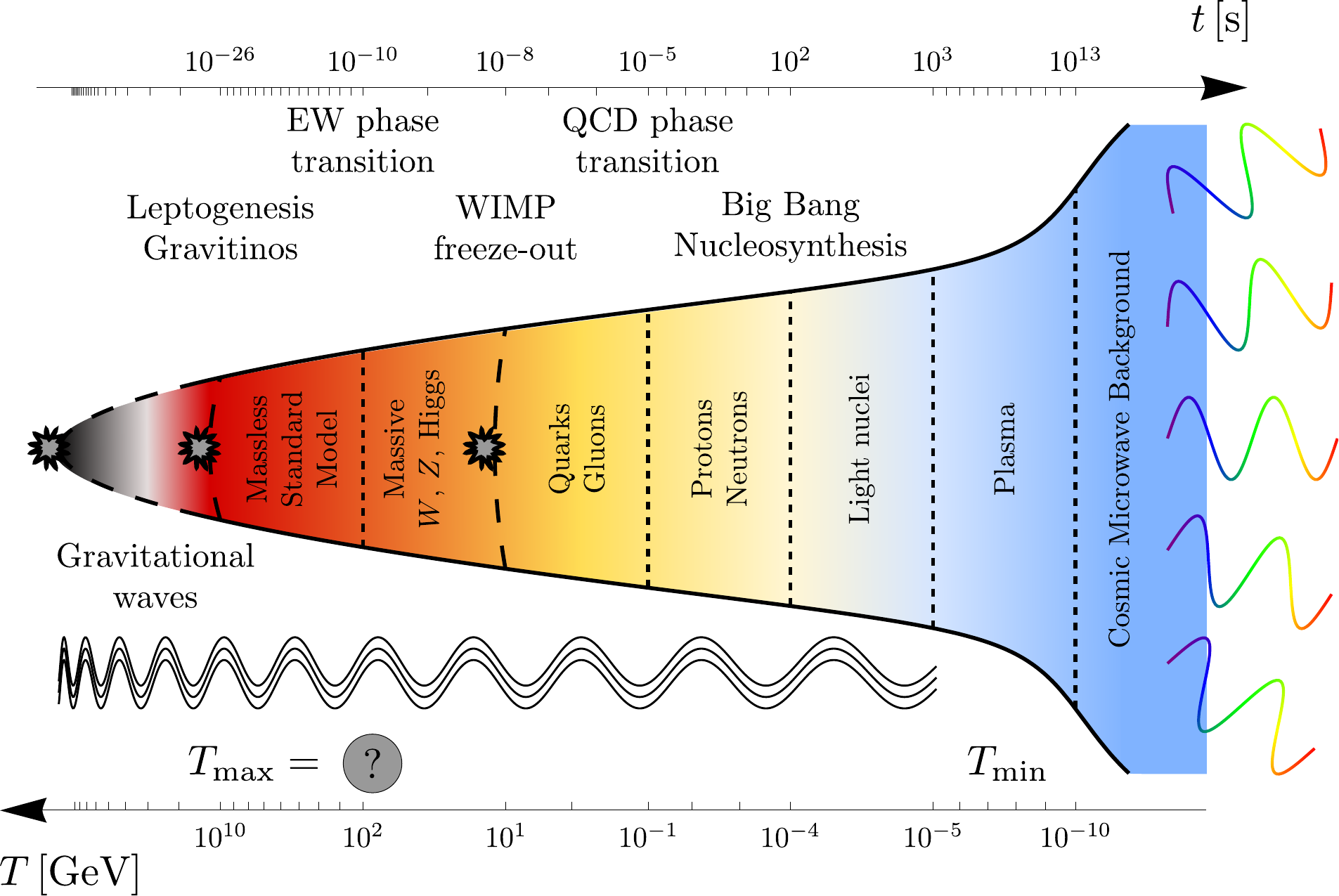}}
\caption{Epochs of the hot early universe, their cosmic time
  scales (top) and temperatures (bottom). From Ref.~\cite{kai}.}
\label{earlyuniverse}
\end{figure}
\item
$T_R\sim 100~{\rm GeV}$ [$t \sim 10^{-10}~{\rm s}$]\\
The Higgs vacuum expectation value forms, and all Standard Model
particles become massive. Baryon and lepton number changing `sphaleron processes' are
no longer in thermal equilibrium.
\item
$T_R\sim 10^8\ldots 10^{11}~{\rm GeV}$ [$t\sim 10^{-22}\ldots
10^{-28}~{\rm s}$]\\
Baryogenesis via leptogenesis takes place and gravitino dark matter
can be thermally produced. 
\item
$T_R\sim 10^{12}~{\rm GeV}$ [$t\sim 10^{-30}\ {\rm s}$]
In supersymmetric theories with extra dimensions one expects that the
present `vacuum' of the universe is metastable \cite{Kachru:2003aw}.  To avoid a
rapid transition to a supersymmetric flat higher-dimensional ground state, 
the reheating temperature cannot exceed a `maximal' reheating temperature
$T_R^{\rm max}$. For gravitino masses $\mG  = {\cal O}(\mathrm{TeV})$, one
estimates in string theories $T_R^{\rm max} \sim  10^{12}~{\rm GeV}$ 
\cite{Buchmuller:2004xr}.
\end{itemize}

The epochs of the hot early universe described above are illustrated in
Fig.~\ref{earlyuniverse}. The CMB provides us with detailed
information about the final stage of the hot phase. Hence, the
reheating temperature must have exceeded $\sim 0.1~{\rm eV}$. 
Furthermore, the success of primordial nucleosynthesis suggests that the
reheating temperature has reached $\sim 10~{\rm MeV}$. Here our
present knowledge ends. Progress will crucially depend on how much
the various possibilities for baryogenesis and dark matter candidates
can be narrowed down. The following four examples illustrate the
impact on the reheating temperature $T_R$. It would be most
fascinating to obtain direct information about the beginning of the
hot early universe, which may eventually be achieved by means of
gravitational waves \cite{Nakayama:2008wy}.

\section{Example I: Moduli Decay}

Let us first consider an example \cite{Kitano:2008tk} with a very low reheating temperature,
$T_R \sim 100~\mathrm{MeV}$, just above the temperature required by
BBN. The theoretical framework is a
supersymmetric extension of the Standard Model with a heavy `modulus
field', which is typical for string compactifications. The initial
energy density of the universe is dominated by coherent oscillations of the
modulus field, with an equation of state corresponding to nonrelativistic matter.
 
The modulus superfield  $\Phi=(\phi,\widetilde{\phi},F_\phi)$ is
assumed to  couple to matter fields via a specific nonrenormalizable interaction
in the superpotential, suppressed by an inverse power of the Planck
mass $\Mp$,
\bea\label{coupling}
 W\,\supset\, 
 \frac{1}{\Mp}\Phi (U D D)\ ,
\eea
where $U=(\widetilde{u}^c,u^c,F_u)$ and $D=(\widetilde{d}^c,d^c,F_d)$ denote
up- and down-type quark superfields, respectively. The $\phi$ charge
density, the difference between the number densities $\phi$ and
anti-$\phi$ particles, is given by
\bea 
 q_\phi =  n_\phi - n_{\phi^*}    
= i\left(\Dot{\phi}^*\phi-\phi^*\Dot{\phi}\right)\ . 
\eea

The time evolution of modulus field and charge density are determined
by the equations of motion ($H$: Hubble rate,  
$\Gamma_\phi$: decay rate),
\bea
 \Ddot{\phi}+(3H+\Gamma_\phi)\,\Dot{\phi}+\frac{\partial V}{\partial \phi^*}
 &=& 0\ , \\
 \Dot{q}_\phi+3H\,q_\phi &=&
 - i\left(\phi\frac{\partial V}{\partial \phi}-
 \phi^*\frac{\partial V}{\partial \phi^*}\right) \ , 
\eea
where $H$ and $\Gamma_\phi$ denote Hubble parameter and $\phi$-decay
rate, respectively. The scalar potential contains a $\phi$ mass term,
some polynomial $F(| \phi |^2 / \Mp^2 )$ and a power of
$\phi$ that is determined by a discrete symmetry,    
\bea
 V =
 m_\phi^2 | \phi |^2 +  \mG^2 M_{\rm p}^2 F(| \phi |^2 / M_{\rm p}^2 ) 
 + \left(\kappa\,\frac{m_{3/2}^2}{M_{\rm p}^4}\,\phi^6+\text{h.c.}
\right) + \ldots ,
\eea
where $\kappa = \mathcal{O}(1)$ and $m_{3/2}$ is the gravitino mass.

During inflation the modulus field develops an expectation value 
$\phi_\mathrm{ini} \sim M_{\rm p}$ with a phase ${\cal O}(1)$. This
condensate stores a charge density. Integrating the field equations
from the initial state with $H \gg m_\phi$ up to $t_\phi
=m_\phi^{-1}$, one obtains for the charge density  
\bea
 q_\phi( t_\phi )
 ~\sim~
 |\kappa|\,
 \frac{m_{3/2}^2}{2m_\phi\,M_{\rm p}^4}\,\phi_\mathrm{ini}^6 \ .
\eea
The number density of $\phi$ particles is determined by the
energy density of the $\phi$ field,
\bea
n_\phi+n_{\phi^*} \simeq \frac{\rho_\phi}{m_\phi} = 
\frac{1}{m_\phi}\left(m_\phi^2\,|\phi|^2+|\dot\phi|^2\right)\ ,
\eea
from which one obtains for the $\phi$ charge asymmetry:
\bea
 \varepsilon = \frac{n_\phi-n_{\phi^*}}{n_\phi+n_{\phi^*}}
 = \frac{q_\phi}{n_\phi+n_{\phi^*}}
 \sim
 |\kappa|\,\left(\frac{m_{3/2}}{m_\phi}\right)^2 \ .
% \left(\frac{\phi_\mathrm{ini}}{M}\right)^4
\eea
For $\phi < \Mp$, baryon number is approximately conserved, and the $\phi$ 
asymmetry becomes a baryon asymmetry in $\phi$ decays such as  
$\phi^* \to  u d \widetilde{d}$.

The $\phi$ decay width is given by
\bea
 \Gamma_\phi~=~\xi\,\frac{m_\phi^3}{M^2}\ , 
\quad \xi = 10^{-3}\ldots  10^{-2} \ ,
\eea
for a coefficient ${\cal O}(1)$ in Eq.~(\ref{coupling}). Assuming
`instant reheating', one obtains from the definition $H(T_R) = \Gamma_\phi$ the
reheating temperature
\bea\label{TR}
 T_R &\simeq& \left(\frac{90}{\pi^2g_*}\right)^{1/2}\sqrt{\Gamma\Mp}\nonumber\\
&\simeq& 120~\mathrm{MeV}
\left(\frac{\xi}{10^{-2}}\right)^{1/2}
\left(\frac{m_\phi}{1500\,\mathrm{TeV}}\right)^{3/2} \ ,
\eea
where we have used $\sqrt{\pi^2g_*/90}\simeq1$ for temperatures of
order MeV. Until they decay, $\phi$ particles dominate the energy
density of the universe, which relates their number density just
before the decay to the reheating temperature,
\bea
 m_\phi (n_\phi+n_{\phi^*}) \simeq \frac{\pi^2}{30} g_* T_R^4\ .
\eea
Using Eq.~(\ref{TR}), one then obtains the baryon asymmetry in terms
of $\phi$ asymmetry, $\phi$ mass and reheating temperature,
\bea
 \frac{n_b}{s}
 &\simeq&  \frac{3}{4}\,\varepsilon\,\frac{T_d}{m_\phi} \nonumber \\
 &\sim& 
 10^{-10}\ |\kappa|
\left(\frac{\xi}{10^{-2}}\right)^{1/2}
\left(\frac{m_{3/2}}{50~\mathrm{TeV}}\right)^2
\left(\frac{m_\phi}{1500\,\mathrm{TeV}}\right)^{-3/2} \ . 
\eea
Clearly, for a very heavy gravitino, as predicted by anomaly mediation
\cite{Randall:1998uk,Giudice:1998xp}, and an even heavier modulus
field $\phi$, the observed baryon asymmetry can be generated in $\phi$
decays.  

\begin{figure}[t]
\begin{center}
\includegraphics[width=7cm,height=7cm]{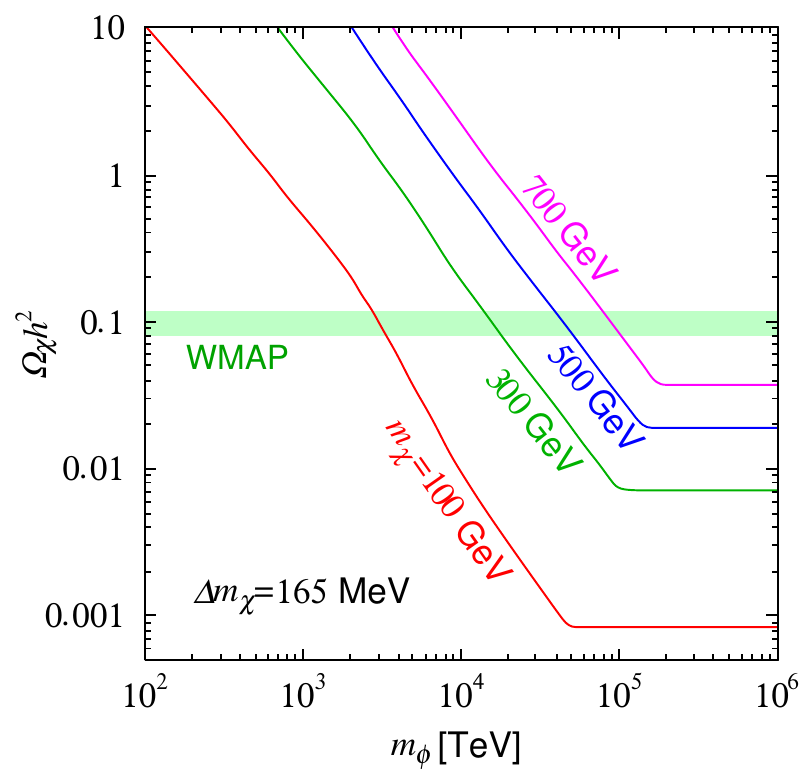}
\end{center}
\caption{$\Omega_\chi h^2$ as function of the modulus mass $m_\phi$
  for various wino masses $m_\chi$.  From Ref.~\cite{Kitano:2008tk}}
\end{figure}

The lightest supersymmetric particle (LSP) is a natural dark matter
candidate. It can be a higgsino or wino, as predicted in anomaly
mediation. Approximately one LSP is produced per $\phi$ decay.
The LSP density is reduced by pair annihilation, and solving a set of
Boltzmann equations leads to the prediction for the dark matter
abundance \cite{Kitano:2008tk}.
\bea
 \Omega_\chi h^2 \simeq 
 0.1
\left(\frac{3\times10^{-3}}{m_\chi^2\langle\sigma v\rangle}\right)
\left(\frac{10^{-2}}{\xi}\right)^{1/2}
\left(\frac{m_\chi}{100\mathrm{GeV}}\right)^3
\left(\frac{m_\phi}{1500\mathrm{TeV}}\right)^{-3/2} .
\eea
In anomaly mediation, one has for the wino LSP,
$m_\chi/m_{3/2}\sim g_2^2/(16\pi^2)$.  In the ratio of dark matter
abundance and baryon asymmetry the dependence on the $\phi$ mass drops
out, and one obtains
 \bea
 \frac{\Omega_\chi}{\Omega_b}
 ~\sim~|\kappa|^{-1}\times
 10^{-2}\times\frac{m_\chi}{m_\mathrm{nucleon}}\ .
\eea
For $m_\chi = {\cal O}(100~\mathrm{GeV})$ the observed ratio  
$\Omega_\mathrm{CDM}/\Omega_b\simeq5$ is easily accomodated.
The observed dark matter abundance imposes a constraint on LSP and
modulus masses, which is illustrated in Fig.~2.

The example of modulus decay nicely illustrates that a reheating
temperature as small as $T_R \sim 100~\mathrm{MeV}$ is sufficient for
a consistent picture of primordial nucleosynthesis, baryogenesis and
dark matter. On the other hand, the predictive power of the model is
rather limited. Two observables, $\Omega_b$ and $\Omega_\mathrm{DM}$
are related to four new parameters, $m_{3/2}$, $m_{\phi}$, $\kappa$
and $\xi$. In addition, the initial value $\phi_{\mathrm{ini}}$
of the modulus field has to be postulated. The model then yields the composition 
of the primordial plasma at a temperature $T_R \sim 100~\mathrm{MeV}$.
It would be very interesting to identify further comological
predictions of the model.

\section{Example II: Electroweak Baryogenesis}

\subsection{The High-Temperature Phase of the Standard Model}

Most mechanisms of baryogenesis make use of some noneqilibrium process
in the hot early universe, such as the decay of heavy particles 
or cosmological phase transitions. One then has to satisfy Sakharov's
conditions for particle interations and cosmological evolution \cite{Sakharov:1967dj},
\begin{itemize}
\item baryon number violation,
\item $C$ and $C\!P$ violation,
\item deviation from thermal equilibrium.
\end{itemize}
Even if these conditions are fulfilled, further severe quantitative constraints
must usually be satisfied to obtain the observed matter-antimatter asymmetry.
This is well illustrated by the example presented in Sakharov's
original paper: Superheavy `maximons' with mass $\mathcal{O}(\Mp)$
decay at an  initial temperature $T_i \sim M_\mathrm{P}$ with a $C\!P$
violation related to the $C\!P$ violation in $K^0$-decays, and the violation 
of baryon number leads to a proton lifetime $\tau_p > 10^{50}\ \mathrm{years}$,
much larger than current estimates in grand unified theories. 

\begin{figure}[t]
\begin{center}
\includegraphics[width=9cm,height=5cm]{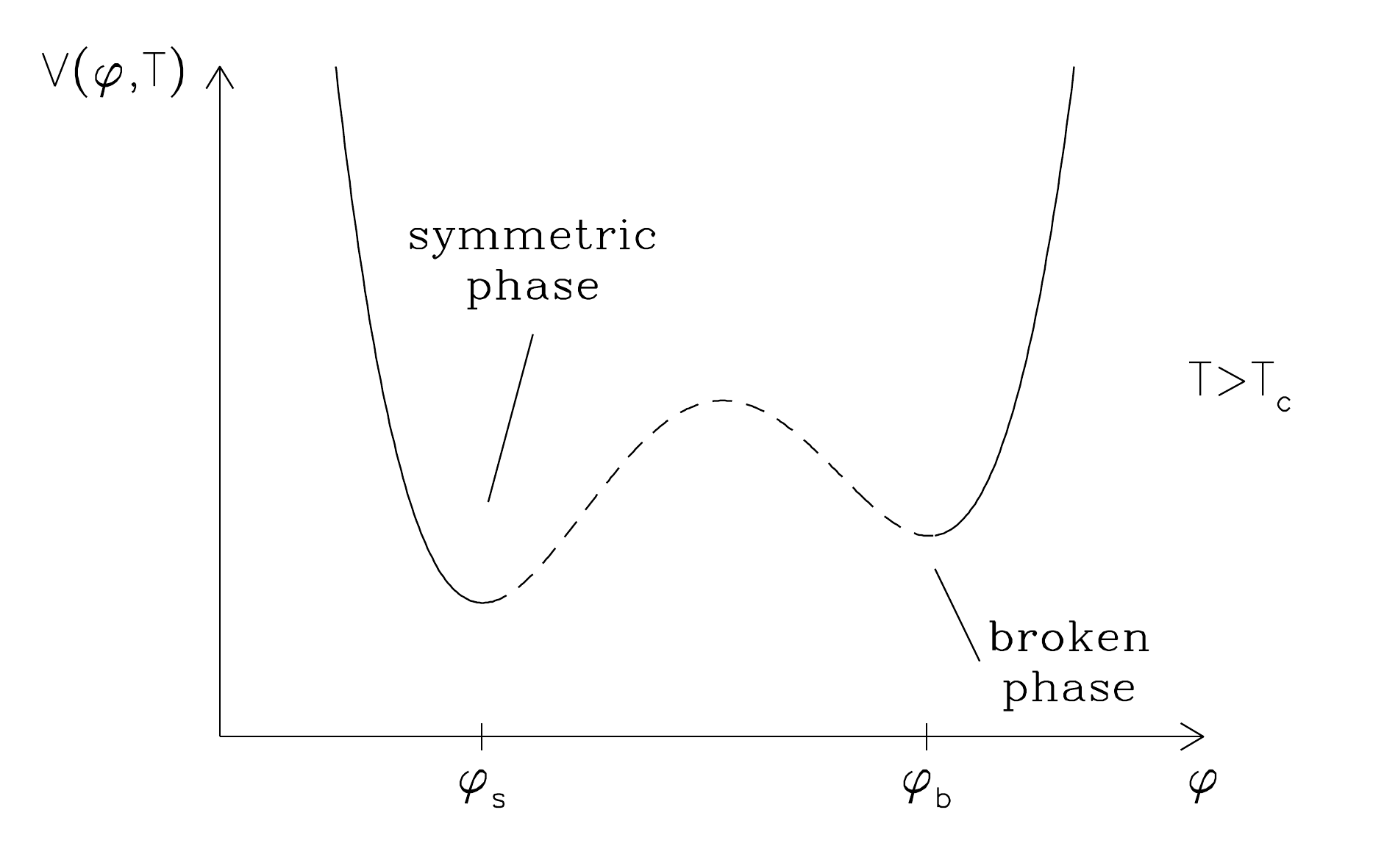}
\end{center}\
\caption{The finite-temperature effective potential of the Higgs field
  above the critical temperature, i.e. $T > T_c$. $\phi_s$ and
  $\phi_b$ correspond to symmetric and Higgs (broken) phase.}  
\end{figure}

The theory of baryogenesis crucially depends on nonperturbative properties
of the standard model, first of all the nature of the electroweak
transition. Depending on the temperature, the symmetric phase or the
broken phase represents the global minimum (cf.~Fig.~3). At the
critical temperature $T_c$ both phases are degenerate. At temperatures
just above (below) $T_c$, a first-order phase transition can occur from the
symmetric (broken) to the broken (symmetric) phase. A measure for the
strength of the transition is the jump in the expectation value of the
Higgs field,
\bea
v_T = \sqrt{\phi_b^{\dagger} \phi_b}\Big|_T - \sqrt{\phi_s^{\dagger}
  \phi_s}\Big|_T \ .  
\eea
\begin{figure}[t]
\begin{center}
\includegraphics[width=6cm,height=5.7cm]{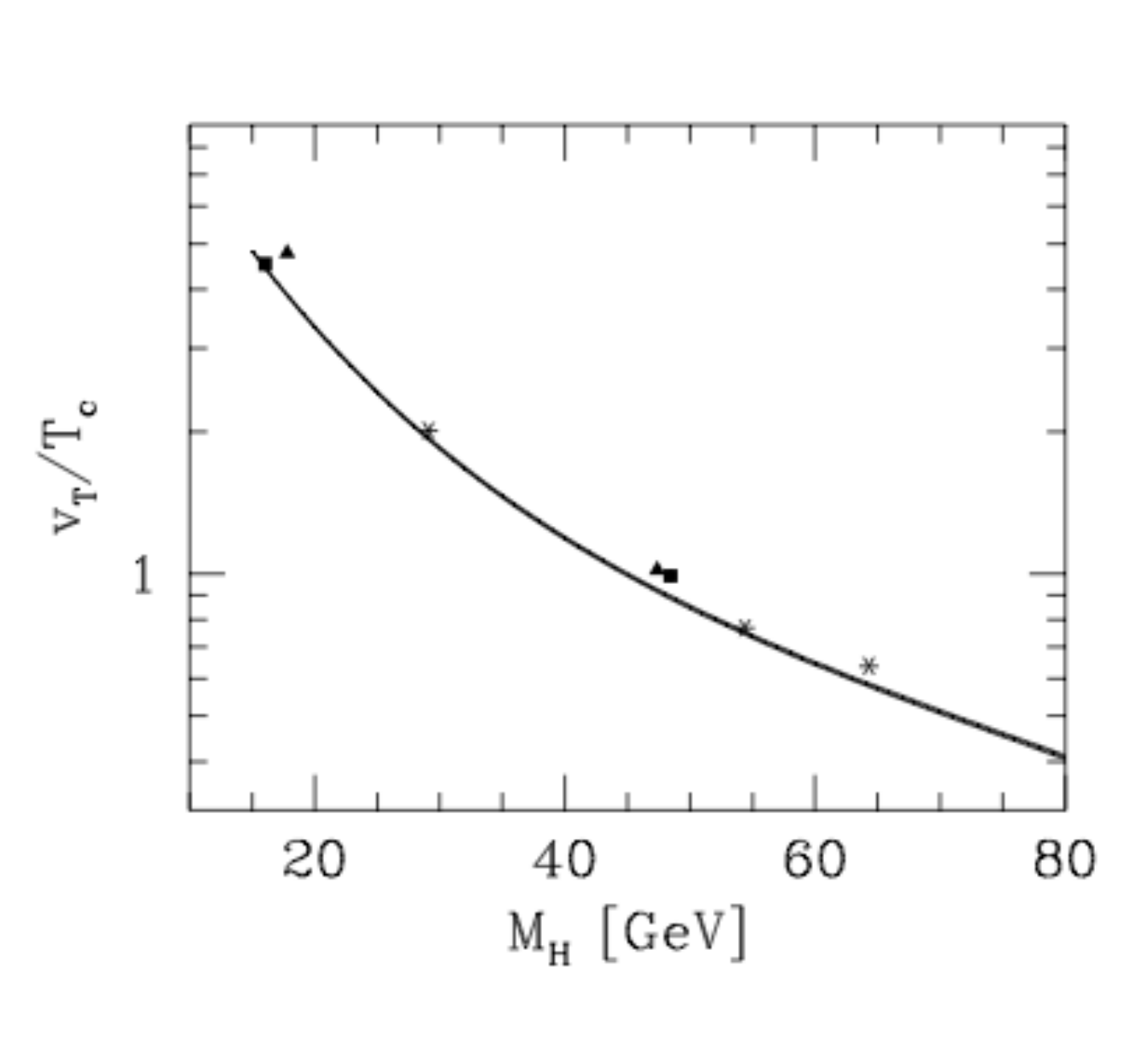}\hspace*{0.7cm}
\includegraphics[width=6cm,height=6.15cm]{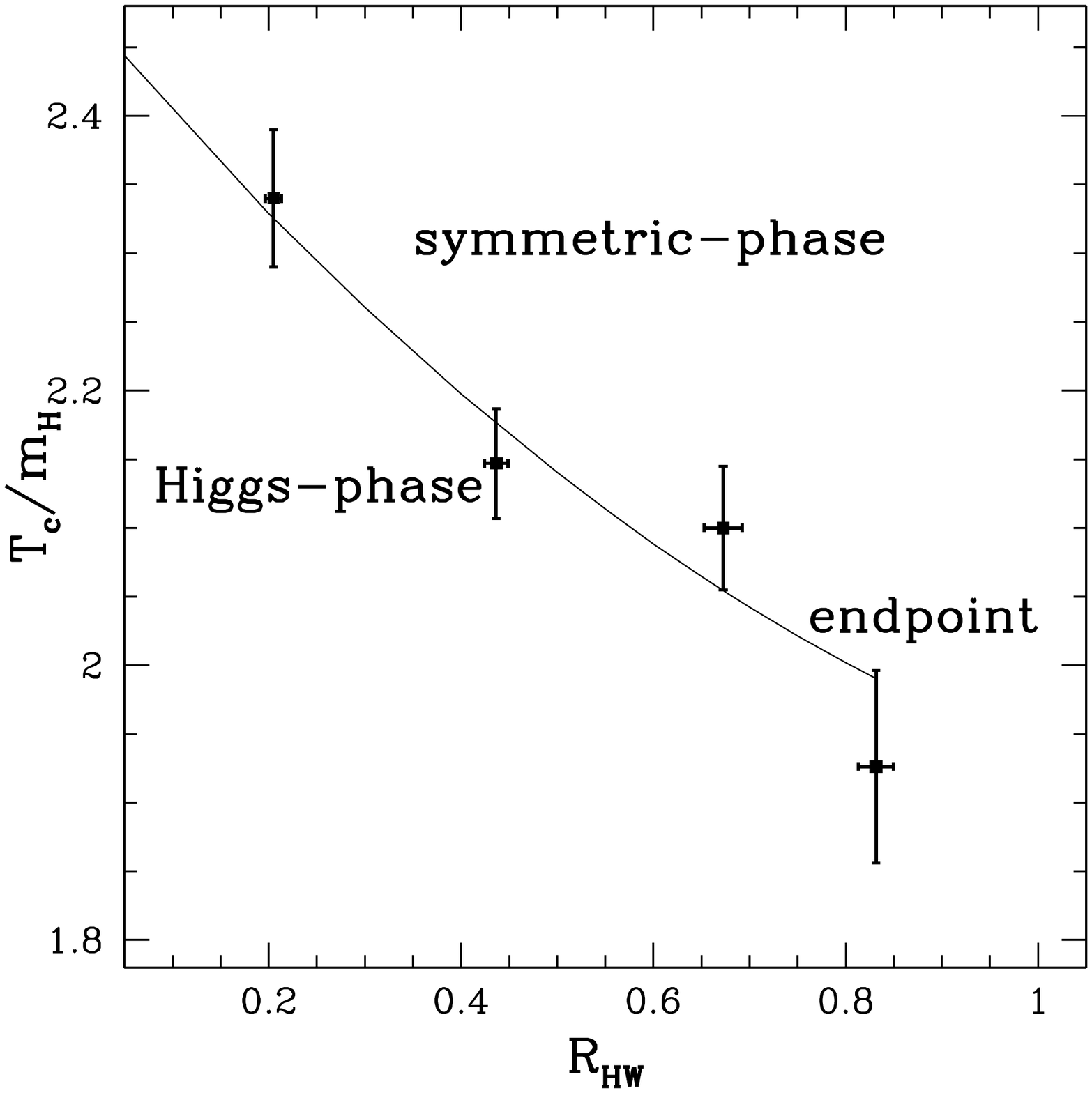}
\end{center}
\caption{(a) (left) Jump of the Higgs expectation value at the
  critical temperature as function of the Higgs mass. Comparison of
  four-dimensional lattice simulations (triangles, squares) \cite{Fodor:1994sj} with
  three-dimensional lattice simulations (stars) \cite{Kajantie:1995kf} and resummed
  perturbation theory \cite{Buchmuller:1995sf}. From Ref.~\cite{Jansen:1995yg}. (b)(right) Critical temperature as
  function of Higgs/W-boson mass ratio $R_{HW}=m_H/m_W$ from
  four-dimensional lattice simulations. From Ref.~\cite{Csikor:1998ew}.}
\end{figure}
A comparison between perturbative and lattice calculations of $v_T$ at
$T=T_c$ is shown in Fig.~4a. There is remarkable agreement between
lattice simulations for the full
four-dimensional theory at finite temperature, for the effective
three-dimensional theory corresponding to the high-temperature limit,
and resummed perturbation theory.

The perturbative result for $v_T$ shows a smooth decrease for Higgs
masses up to $80~\mathrm{GeV}$. This behaviour, however, is not
correct since for large Higgs masses nonperturbative effects become
important, which turn the first-order phase transition into a smooth
crossover for Higgs masses larger than $m_H^c = \mathcal{O}(m_W)$. This
behaviour has first been demonstrated by solving gap equations for the
Higgs model \cite{Buchmuller:1994qy} and then by three-dimensional lattice
simulations \cite{Kajantie:1996mn}. The result of four-dimensional
lattice simulations \cite{Csikor:1998ew} is displayed in Fig.~4b,  
which gives $T_c/m_H$ as
function of the Higgs mass in units of the W-boson mass,
$R_{HW}=m_H/m_W$. The line of first-order phase
transitions for small Higgs masses has an endpoint corresponding to a
second order phase transition. The corresponding critical Higgs mass is 
$m_H^c = 72.1 \pm 1.4~\mathrm{GeV}$ \cite{Fodor:1999at}. For larger Higgs masses
the electroweak transition is a smooth crossover. Critical temperature
and critical Higgs mass can be estimated by requiring that the
perturbative vector boson mass $m = gv_T$ is equal to the
nonperturbative finite-temperature magnetic mass, obtained by solving
gap equations, $m_{SM}=Cg^2 T$, with $C\simeq 0.35$ 
\cite{Eberlein:1998mb,Bieletzki:2012rd}. This yields
for the critical Higgs mass \cite{Buchmuller:1996pp}
\bea
m_H^c = \left({3\over 4\pi C}\right)^{1/2} m_W \simeq 74~\mbox{GeV}\ .
\eea
The critical Higgs mass is far below the mass of $126~\mathrm{GeV}$
of the Higgs-like boson recently discovered at the LHC \cite{:2012gk}. If this
boson is indeed the Higgs particle of the Standard Model, then we know
that there has been no departure from themal equilibrium during the
cosmological electroweak transition. 

The second crucial nonperturbative aspect of baryogenesis is the connection 
between baryon number and lepton number in the high-temperature, symmetric 
phase of the Standard Model. Due to the chiral nature of the weak interactions 
$B$ and $L$ are not conserved \cite{'tHooft:1976up}. At zero temperature this has no 
observable effect due to the smallness of the weak coupling. However, as the 
temperature reaches the critical temperature $T_c$ of the electroweak phase 
transition, $B$ and $L$ violating processes come into thermal 
equilibrium \cite{Kuzmin:1985mm}. The rate of these processes is
related to the free energy of sphaleron-type field configurations which carry
topological charge. In the standard model they lead to an effective
interaction of all left-handed fermions \cite{'tHooft:1976up} (cf.~Fig.~5), 
\bea
O_{B+L} = \prod_i \left(q_{Li} q_{Li} q_{Li} l_{Li}\right)\; ,
\eea
which violates baryon and lepton number by three units, 
\bea
    \Delta B = \Delta L = 3\;. \label{sphal1}
\eea
\begin{figure}[t]
\begin{center}
\includegraphics[height=5cm]{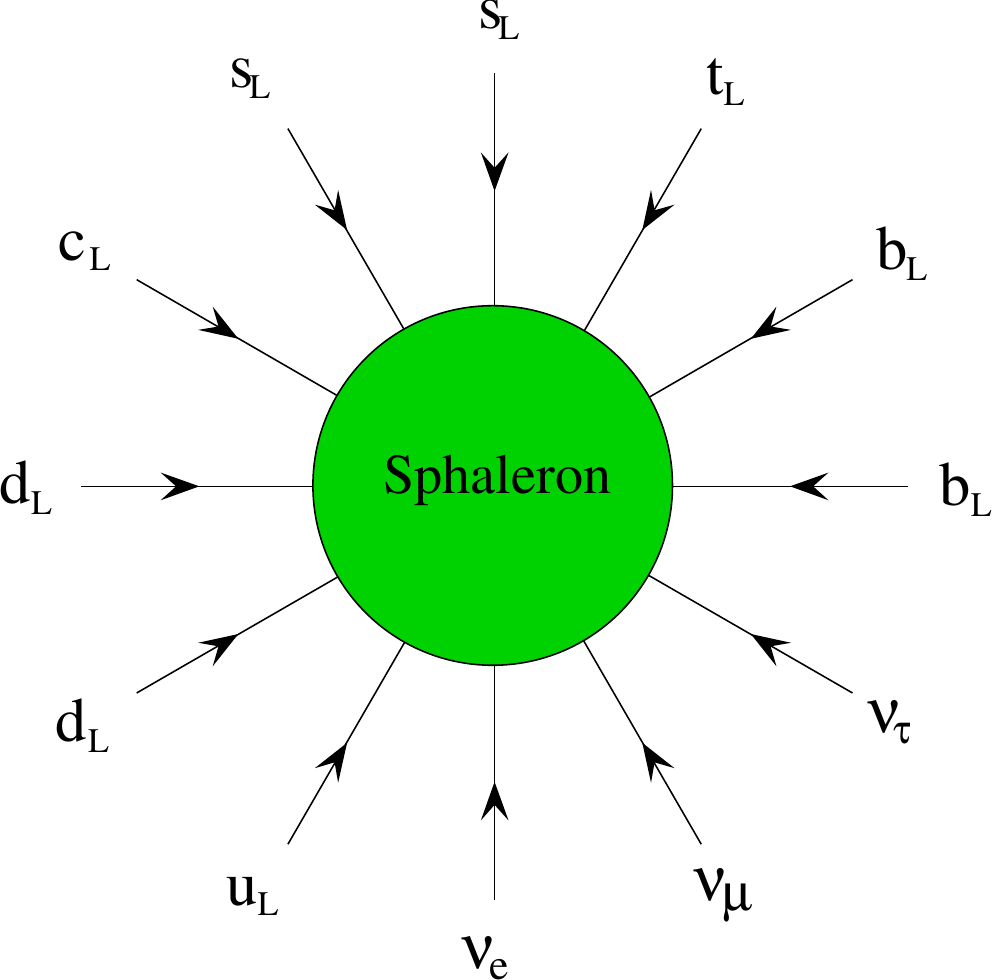}
\end{center}
\caption{One of the 12-fermion processes which are in thermal equilibrium
in the high-temperature phase of the standard model.
\label{fig:sphal}}
\end{figure}
The sphaleron transition rate in the symmetric high-temperature phase
has been evaluated by combining an analytical resummation with numerical
lattice techniques \cite{Bodeker:1999gx}. The result is, in accord with previous 
estimates, that $B$ and $L$ violating processes are in thermal equilibrium for 
temperatures in the range
\bea
T_{EW} \sim 100\ \mbox{GeV} < T < T_{SPH} \sim 10^{12}\ \mbox{GeV}\;.
\eea
Although uncontroversial among theorists, it has to be stressed that this
important phenomenon has so far not been experimentally tested! It is
therefore very interesting that the corresponding phenomenon of
chirality changing processes in strong interactions might be
observable in heavy ion collisions at the LHC 
\cite{Kharzeev:2007jp,Kalaydzhyan:2011vx}.

Sphaleron processes relate baryon and lepton number and therefore
strongly affect the generation of the cosmological baryon asymmetry. 
Analyzing the chemical potentials of quarks and leptons in thermal
equilibrium \cite{Harvey:1990qw}, one obtains an important relation between the asymmetries
in $B$-,  $L$- and $B$-$L$-number,
\begin{equation}
\langle B\rangle_T = c_S \langle B-L\rangle_T 
= {c_S\over c_S-1} \langle L\rangle_T\ ,
\end{equation}
where $c_S = {\cal O}(1)$. In the Standard Model one has $c_s= 28/79$. 

This relation suggests that lepton number violation can explain
the cosmological baryon asymmetry. However, lepton number violation
 can only be weak at late
times, since otherwise any baryon asymmetry would be washed out. 
The interplay of these 
conflicting conditions leads to important contraints on neutrino properties,
and on extensions of the Standard Model in general. Because of the
sphaleron processes, lepton number violation can replace baryon number
violation in  Sakharov's conditions for baryogenesis.

\subsection{Composite Higgs Model}

Baryogenesis requires departure from thermal equilibrium. As discussed
in the previous section, due to the large Higgs mass the electroweak
transition in the Standard Model does not provide the necessary noneqilibrium 
for electroweak baryogenesis. However, in extensions of the
Standard Model with a strongly interacting Higgs sector, sufficiently
strong first-order electroweak phase transitions \cite{Espinosa:2011ax}
and electroweak baryogenesis are possible \cite{Espinosa:2011eu}.

As an example, consider a strongly interacting theory where a global
$SO(6)$ symmetry is spontaneously broken to the subgroup $SO(5)$,
such that the Higgs doublet together with an additional singlet $s$
arise as pseudo-Goldstone bosons. In unitary gauge, the
finite-temperature potential for the Higgs field $h$ and the singlet
$s$ can be written as \cite{Espinosa:2011eu},
\bea
V(h,s,T) &=& \frac{\lambda_h}4 \left[ h^2 - v_c^2 + 
\frac{v_c^2}{w_c^2} s^2 \right]^2 + \frac{\kappa}4 \,  s^2 h^2 \\
&& + \frac{1}{2}(T^2 - T_c^2) (c_h \, h^2 + c_s \, s^2 )\ ,
\eea
\begin{table}[t]
\begin{tabular}{|c|c|c|c|c|c|c|}
\hline
& $m_h$ & $m_s$ & $v_c$  & $f/b$ & $L_w v_c$ & $v_c/T_c$\\
\hline
\hline
S1 & $120$ GeV & $81$ GeV  & $188$ GeV & $1.88$ TeV & $7.1$ & $2.0$\\
\hline
S2 & $140$ GeV & $139.2$ GeV  & $177.8$ GeV  & $1.185$ TeV & $3.5$ & $1.5$\\
\hline
\end{tabular}
\caption{Two numerical examples of models with viable eletroweak
  baryogenesis; $m_h$ and $m_s$ are the Higgs and singlet masses,
  respectively. From Ref.~\cite{Espinosa:2011eu}.}
\end{table}
where $v_c$, $w_c$, $\kappa$, $c_h$ and $c_s$ are parameters of the
model. For realistic Higgs masses, a sufficiently strong first-order
phase transition is achieved, satisfying the necessary condition for
the jump in the Higgs field $v_c/T_c > 1$ (cf.~Table 1).
\begin{figure}[b]
\begin{center}
\includegraphics[width=8cm]{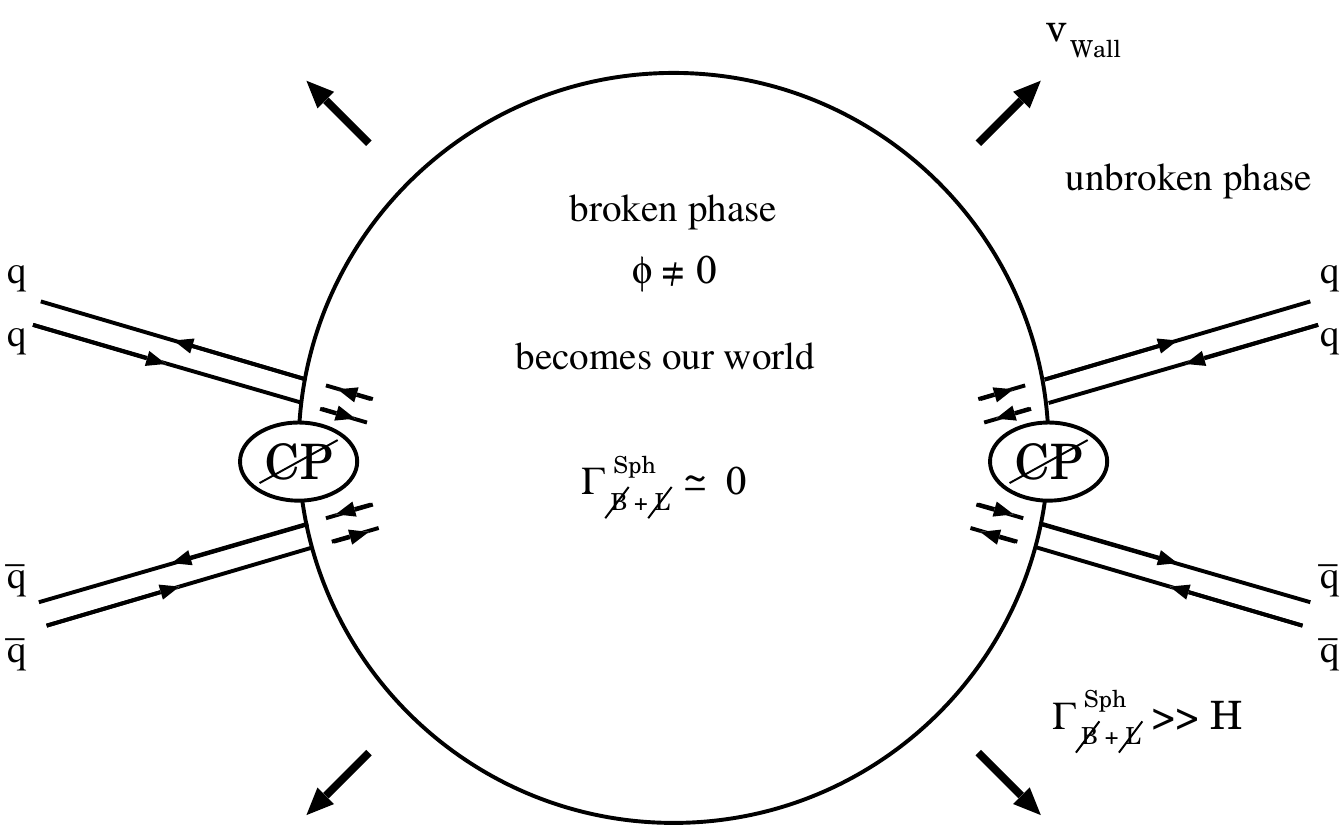}
\end{center}
\caption{Sketch of nonlocal electroweak baryogenesis. From 
Ref.~\cite{Bernreuther:2002uj}.}
\end{figure}

The cosmological first-order phase transition proceeds via nucleation
and growth of bubbles \cite{Bernreuther:2002uj}. This provides the necessary
departure from thermal equilibrium. $C\!P$-violating reflections and
transmissions at the bubble surface then generate an asymmetry in
baryon number (cf.~Fig.~6) , and for a sufficiently strong phase
transition this asymmetry is frozen in the true vacuum inside the bubble

In the considered model, $C\!P$-violating couplings of top-quarks to
the two Higgs bosons $H$ and $s$ are responsible for the generation of
a baryon asymmetry,
\bea
\mathcal{L}_{tHS} = \frac{s}{f} H \bar{Q}_3(a+ib\gamma_5) t + {\mathrm h.c.}\ .
\eea
Detailed calculations show that the observed baryon asymmetry can be
explained for a sufficiently strong coupling. 

The $C\!P$-violating top-quark couplings induce via higher-order loops
electric dipole moments for neutron and electron (cf.~Fig.~7) , which
are severely constrained, $d_e/e < 1.05 \times 10^{-27}\textrm{cm}$ and
$d_n/e < 2.9 \times 10^{-26}\textrm{cm}$ \cite{Beringer:1900zz}.
\begin{figure}[t]
\begin{center}
\includegraphics[width=4.5cm]{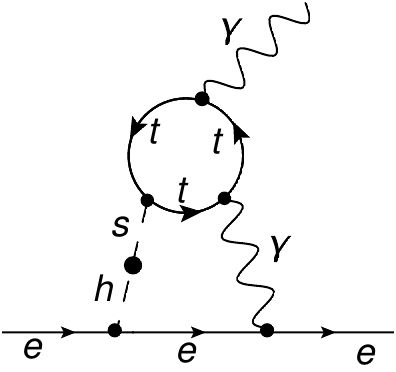}
\end{center}
\caption{Dominant two-loop contribution to electron electric dipole
  moment, induced by $C\!P$ violating top-quark couplings. From 
Ref.~\cite{Espinosa:2011eu}.}
\end{figure}
Furthermore, stringent constraints from electroweak precision
observables have to be satisfied. A consistent picture can be obtained
for Higgs masses in the range from $100~\mathrm{GeV}$ to
$150~\mathrm{GeV}$, in agreement with the evidence for a Higgs-like
particle at the LHC. Note that the considered model in its simplest
form does not have a dark matter candidate. This can be changed by
extending the model to include another, `inert' Higgs doublet. The
required reheating temperature for baryogenesis is the electroweak
scale, $T_R \sim T_{EW} \sim 100~\mathrm{GeV}$.
 
Electroweak baryogenesis in the minimal supersymmetric standard model (MSSM)
has been a very popular scenario after it became clear that baryogenesis
was impossible in the non-supersymmetric Standard Model. It turns out,
however,  that due to recent results from the LHC rather extreme stop
masses are required to make baryogenesis possible, 
$m_{\tilde{t}_R} \lesssim 110~\mathrm{GeV}$ with 
$m_{\tilde{t}_L} \gtrsim 50~\mathrm{TeV}$ \cite{Carena:2012np}, and it
appears likely that electroweak baryogenesis will
soon be ruled out also in the MSSM.

\section{Example III: Baryogenesis in the $\nu$MSM}

At temperatures around the electroweak scale, sphaleron processes come into thermal
equilibrium and baryon number violation can therefore be replaced by
lepton number violation.   This occurs in the Standard Model supplemented
by Majorana (sterile) neutrinos. Baryogenesis can then take place via
leptogenesis \cite{Fukugita:1986hr}. It is remarkable that for small
neutrino masses of order GeV or  keV, such a model ($\nu$MSM scenario)
can account not only for neutrino oscillations, but also for
baryogenesis and dark matter \cite{Asaka:2005an}.

The starting point is the familar Standard Model Lagrangian extended
by Dirac and Majorana mass terms,
\bea
\mathcal{L}_{\nu MSM} &=&\mathcal{L}_{SM} - \bar{L}_{L}F\nu_{R}\tilde{\Phi} -\bar{\nu}_{R}F^{\dagger}L_L\tilde{\Phi}^{\dagger} \nonumber\\ 
&&\quad - \frac{1}{2}(\bar{\nu_R^c}M_{M}\nu_{R} 
	+\bar{\nu}_{R}M_{M}^{\dagger}\nu^c_{R})\ .
\eea
Here $M_M$ is the Majorana mass matrix of the right-handed neutrinos
$\nu_R$, and the Higgs expectation value $\langle \Phi \rangle = v$
generates the Dirac neutrino mass matrix $m_D = F v$. The light and
heavy neutrino mass eigenstates $\nu_i$ and $N_I$ have masses $m_i$
and $M_I$, respectively. A crucial quantity for phenomenology is
the  active-sterile mixing matrix  $\theta=m_D M_M^{-1}$, with 
$U^2 = {\rm tr}(\theta^\dagger\theta)$. 

Recently, the scenario has
been studied in detail quantitatively \cite{Canetti:2012vf}. 
The lightest sterile neutrino $N_1$ provides dark matter, with a mass in
the range $1~\mathrm{keV} < M_1 \lesssim 50~\mathrm{keV}$, and tiny
mixings,  $10^{-13}\lesssim\sin^2(2\theta_{\alpha1})\lesssim 10^{-7}$, 
constrained by X-ray observations. The allowed and excluded parameter
regions are shown in Fig.~8. Along the solid lines the model
reproduces the observed value of $\Omega_{\rm DM}$ for the indicated
chemical potentials. 

Following Ref.~\cite{Akhmedov:1998qx}, baryogenesis is achieved by  
$C\!P$-violating oscillations of $N_{2}$ and
$N_3$, which are thermally produced at temperature $T \gtrsim T_{EW}
\sim 140~\mathrm{GeV}$ (assuming a Higgs mass $m_H = 126~\mathrm{GeV}$).
The time evolution of the $N_{2,3}$ density matrices $\rho_N$ and
$\rho_{\bar{N}}$  and the lepton chemical potentials
$\mu_{\alpha}$ are described by the kinetic equations \cite{Canetti:2012vf}
\bea
i \frac{d\rho_{N}}{d T}&=&[H, \rho_{N}]-\frac{i}{2}\{\Gamma_N, \rho_{N} 
- \rho^{eq}\} +\frac{i}{2} \mu_\alpha{\tilde\Gamma^\alpha_N}~,\\
i \frac{d\rho_{\bar{N}}}{d T}&=& [H^*, \rho_{\bar{N}}]-\frac{i}{2}
\{\Gamma^*_N, \rho_{\bar{N}} - \rho^{eq}\} -
\frac{i}{2} \mu_\alpha{\tilde\Gamma_N^{\alpha *}}~,\label{kinequ2}\\
i \frac{d\mu_\alpha}{d T}&=&-i\Gamma^\alpha_L \mu_\alpha +
i {\rm tr}\left[{\tilde \Gamma^\alpha_L}(\rho_{N} -\rho^{eq})\right]
\nonumber\\ 
&&-i {\rm tr}\left[{\tilde \Gamma^{\alpha*}_L}(\rho_{\bar{N}} 
 -\rho^{eq})\right]\ .
\eea
Here $\rho^{eq}$ is the equilibrium density matrix, $H$ is the
dispersive part of the finite temperature effective Hamiltonian for
the $N_I$ and $\Gamma_N$, $\Gamma_L^{\alpha}$ and
$\tilde{\Gamma}_L^{\alpha}$ are rates accounting for different dissipative
effects. The equations describe thermal sterile neutrino production,
oscillations, freeze-out and decay. 
\begin{figure}[t]
\begin{center}
\includegraphics[width=9cm]{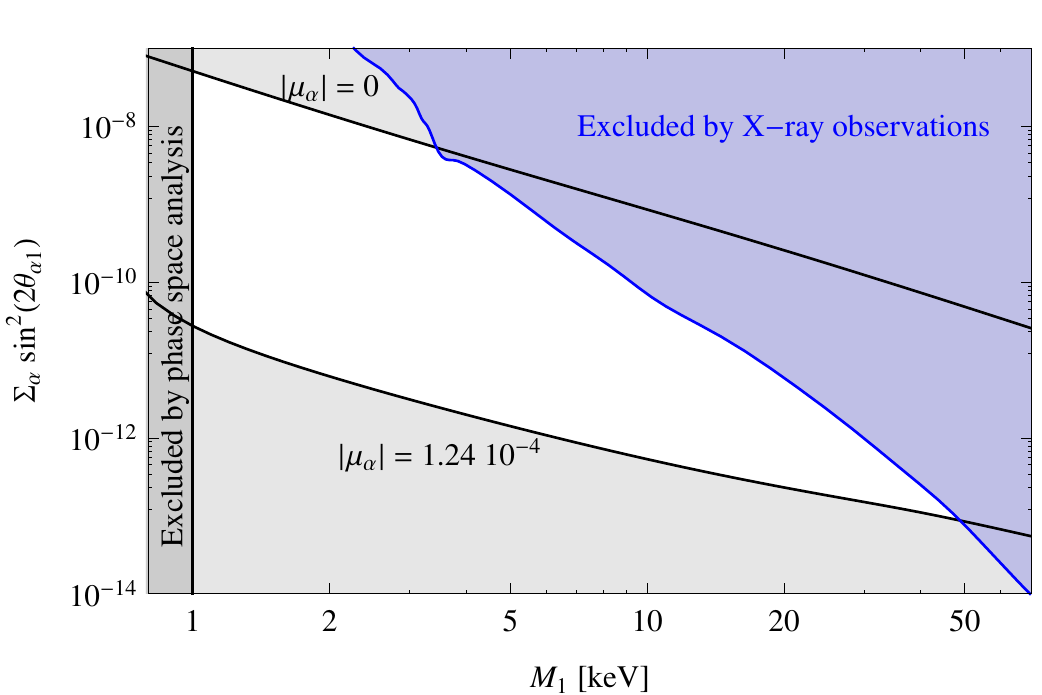}
\caption{Constraints on $N_1$ mass and mixing. The blue region is
  excluded by X-ray observations, the dark gray region ($M_1 <
  1~\mathrm{keV}$) by the Tremaine-Gunn bound. On the solid lines the
  model reproduces the observed value of $\Omega_{\rm DM}$ for the
  indicated chemical potentials. From Ref.~\cite{Canetti:2012vf}.}
\end{center}
\end{figure}
\begin{figure}[t]
\begin{center}
\includegraphics[width=9cm]{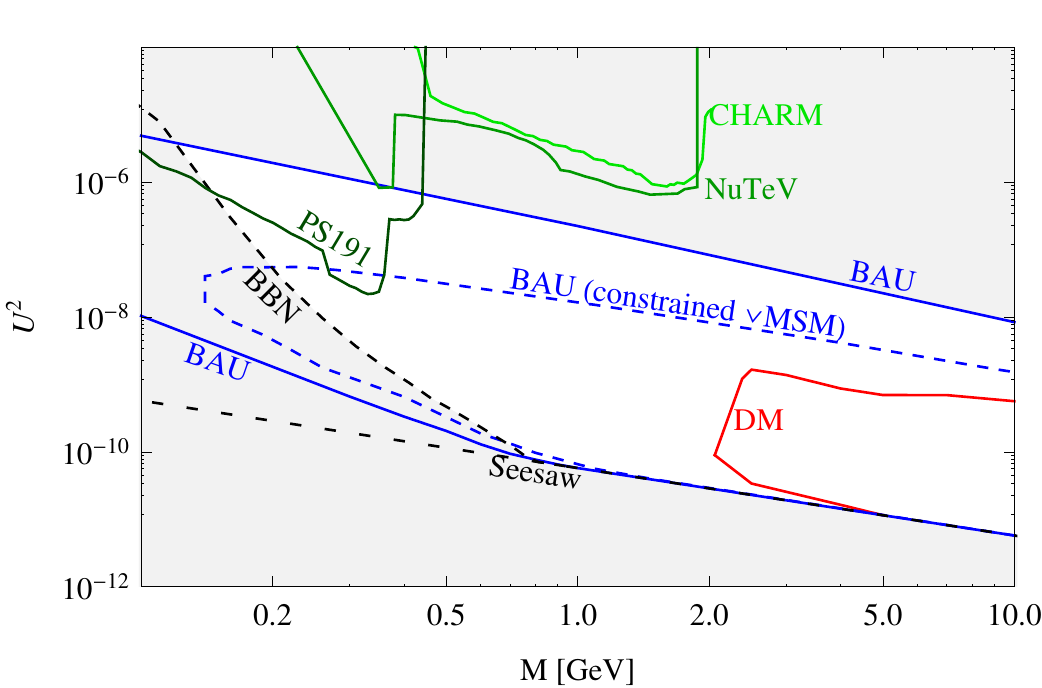}
	\caption{Experimental and cosmological constraints on mixing
          $U^2 = {\rm tr}(\theta^\dagger\theta)$ and $N_{2,3}$ masses
          $M_{2,3} \simeq M$.  From Ref.~\cite{Canetti:2012vf}.}
	\label{fig:example}
\end{center}
\end{figure}

To obtain the right amount of baryon asymmetry, resonant enhancement
of $C\!P$ violation is needed, with a very high mass degeneracy
of the sterile neutrinos, $|M_2-M_3|/|M_2+M_3| \sim 10^{-11}$. The
required lepton chemical potential is
\bea
\mu_\alpha \sim 10^{-10} \quad \text{\rm at} \quad T\sim T_{EW}\ .
\eea
At temperatures below $T_{EW}$, the sphaleron processes are
ineffective, so that a change of the lepton chemical potential does
not influence the baryon asymmetry anymore. Now larger lepton chemical
potentials are needed to generate the observed amount of dark matter,  
\bea
|\mu_\alpha|\gtrsim 8\cdot 10^{-6} \quad \text{rm at} \quad 
T\sim 100~\mathrm{ MeV} \ .
\eea
The observed dark matter abundance $\Omega_{\rm DM}$ also restricts the
$N_{2,3}$ masses to lie in the range 2-10 GeV (cf.~Fig.~9). Inflation can be
incorporated by adding a light dilaton field. 

The $\nu$MSM scenario requires a reheating temperature $T_R = 
\mathcal{O}(100\ \mathrm{GeV})$. It is indeed a minimal model and can
be verified or falsified in the near future by astrophysical
observations and collider experiments.

\section{Example IV: Thermal Leptogenesis}

In the previous section we saw that the Standard Model supplemented by
three right-handed neutrinos can explain baryogenesis and dark matter 
for judiciously chosen neutrino masses smaller than the electroweak scale.
On the contrary, in the canonical GUT scenario, the sterile neutrinos
have GUT scale masses, and the smallness of the light neutrino masses
is obtained for Dirac neutrino masses comparable to charged lepton and
quark masses. In this case decays of $N_1$, the lightest of the sterile neutrinos,
generate the baryon asymmetry \cite{Fukugita:1986hr}. The mass $M_1$ of
$N_1$ is much larger than the electroweak scale, and since the $N_1$ abundance is
thermally produced, also the reheating temperature must be much larger
than the electroweak scale.  Clearly, dark matter is now independent
of neutrino physics.
  
For third generation
Yukawa couplings ${\cal O}(1)$, as in some SO(10) GUT models, one
obtains the heavy and light neutrino masses,
\bea
M_3 \sim \Lambda_{\rm GUT} \sim 10^{15}~{\rm GeV} \ , \quad 
m_3 \sim {v^2\over M_3} \sim 0.01~{\rm eV}\ .
\eea
Remarkably, the light neutrino mass $m_3$ is comparable to
$(\Delta m^2_{atm})^{1/2} \equiv m_{\rm atm} \simeq 0.05$~eV, as measured in
atmospheric $\nu$-oscillations. This supports the hypothesis that
neutrino physics probes the mass scale of grand unification.

In the case of hierarchical heavy Majorana neutrinos, the interactions of
$N\equiv N_1$, the lightest of them, with the 
Higgs doublet $\phi$ and the lepton doublets $l_{Li}$ are described by
the effective Lagrangian \cite{Buchmuller:2000nd}
\bea
\mathcal{L} &=& \overline{l}_{Li} \widetilde{\phi}\lambda^*_{i1} N + 
      N^T \lambda_{i1} C l_{Li}\phi - \frac{1}{2}M N^T C N \nonumber\\
&& + \frac{1}{2}\eta_{ij} l_{Li}^T\phi\ Cl_{Lj}\phi
   + \frac{1}{2}\eta^*_{ij} \overline{l}_{Li}\widetilde{\phi}\ C 
    \overline{l}_{Lj}^T\widetilde{\phi}\ ,
\eea
where $\widetilde{\phi} = i\sigma_2\phi^*$ and $C$ is the charge conjugation matrix.
The quartic coupling 
\bea
\eta_{ij}=\sum_{k>1}\lambda_{ik}\frac{1}{M_k}\lambda^T_{kj}\ 
\eea
is obtained after integrating out heavy Majorana neutrinos $N_{k>1}$ with
$M_{k>1} \gg M_1 \equiv M$.  Note that in quantum corrections the
coupling  $\eta$ takes care of vertex and self-energy contributions.
$N$ has small Yukawa couplings, $\lambda_{i1} \ll 1$, and its decay
width $\Gamma$ is therefore much smaller than the mass $M$. 

The heavy Majorana neutrinos have no gauge interactions. Hence, in the
early universe, they can easily be out of thermal equilibrium. This
makes $N$, the lightest of them, an ideal candidate for
baryogenesis, in accord with Sakharov's condition of departure from
thermal equilibrium. In the simplest form of leptogenesis the $N$ abundance
is produced by thermal processes, which is
therefore called `thermal leptogenesis'. The $C\!P$-violating  
$N$ decays into lepton-Higgs pairs lead to a lepton asymmetry 
$\langle L \rangle_T \neq 0$, which is partially converted to a
baryon asymmetry $\langle B \rangle_T \neq 0$ by the sphaleron processes.
In early work on leptogenesis, it was anticipated that the light
neutrino masses are then required to have masses $m_i < \mathcal{O}(1\mathrm{eV})$ 
\cite{Buchmuller:1996pa}.
After the discovery of atmospheric neutrino oscillations, more
stringent upper bounds on neutrino masses could be derived, and
leptogenesis became increasingly popular.

The generated baryon asymmetry is proportional to the $C\!P$ 
asymmetry in $N_1$ decays. For hierarchical heavy neutrinos it
is given by \cite{Covi:1996wh,Flanz:1994yx,Buchmuller:1997yu}
\bea
\epsilon_1 &=& \frac{\Gamma(N_1\rightarrow l\phi)-
\Gamma(N_1 \rightarrow \bar{l}\bar{\phi})}
{\Gamma(N_1 \rightarrow l\phi)+\Gamma(N_1 \rightarrow
  \bar{l}\bar{\phi})} 
\nonumber \\
&=& -\frac{3}{16\pi}\frac{{\rm Im}(m_D^{\dagger}m_{\nu}m_D)_{11}M_1}
{\left(m_D^{\dagger}m_D\right)_{11}v^2} \ .
\eea
From this expression one obtains the estimate \cite{Buchmuller:1998zf}
\bea
\epsilon_1 
&\sim &\ {3\over 16\pi}{m_3 M_1\over v^2} \label{estimate1} \\
&\sim & 0.1\ {M_1\over M_3} \ . \label{estimate2} 
\eea
Note that the r.h.s. of Eq.~(\ref{estimate1}) is in fact a rigorous
upper bound on the $C\!P$ asymmetry $\epsilon_1$ 
\cite{Davidson:2002qv,Hamaguchi:2001gw}. Using the seesaw formula,
Eq.~(\ref{estimate2}) relates the $C\!P$
asymmetry to the mass hierarchy of the heavy neutrinos. 
For mass hierarchies similar to charged lepton and quark mass
hierarchies, $M_1/M_3 \sim 10^{-4}\ldots 10^{-5}$, as expected in
GUTs,  one then obtains the order-of-magnitude estimate
$\epsilon_1 \sim 10^{-5}\ldots 10^{-6}$.  

The small $C\!P$ asymmetry $\epsilon_1$ in the case of hierarchical
heavy neutrinos implies a small baryon asymmetry,
\bea\label{basym}
\eta_B = {n_B - n_{\bar{B}}\over n_\gamma} = -d\ \epsilon_1\ \kappa_f \
\sim\ 10^{-9}\ldots 10^{-10}\ .
\eea
Here the dilution factor $d \sim 0.01$ accounts for the increase of
the photon number density between leptogenesis and today, and the
efficiency factor $\kappa_f \sim 10^{-2}$ is a consequence of washout
effects due to lepton number changing scatterings in the plasma.

It turns out that for the relevant range of neutrino masses, the final
baryon asymmetry is determined by decays and inverse decays of the
heavy neutrinos \cite{Buchmuller:2004nz}. In the ``one-flavour''
approximation, where one sums over lepton flavours in the final state,
the Boltzmann equations take the simple form
\bea
\frac{dn_{N}}{dt} + 3 H n_{N}
&=& -\left(n_{N}-n_{N}^{eq}\right)\ \Gamma_{N}\ , \label{BE1}\\
\frac{dn_L}{dt} + 3 H n_{L}
&=& -\epsilon_1\left(n_{N}-n_{N}^{eq}\right)\ \Gamma_{N} \label{BE2}\ .
\eea
Here $n_{N}$ ($n_{N}^{eq}$) and  $n_{L}$ ($n_{L}^{eq}$) are the
(equilibrium) number densities\footnote{Note that in Fig.~10 number
densities ${\rm N_{N_1}}$ and ${\rm N_{B-L}}$ are plotted for a
portion of  comoving volume that contains one photon.} of heavy
neutrinos and leptons, respectively. 
\begin{figure}[t]
\begin{center}
\includegraphics[width=8cm]{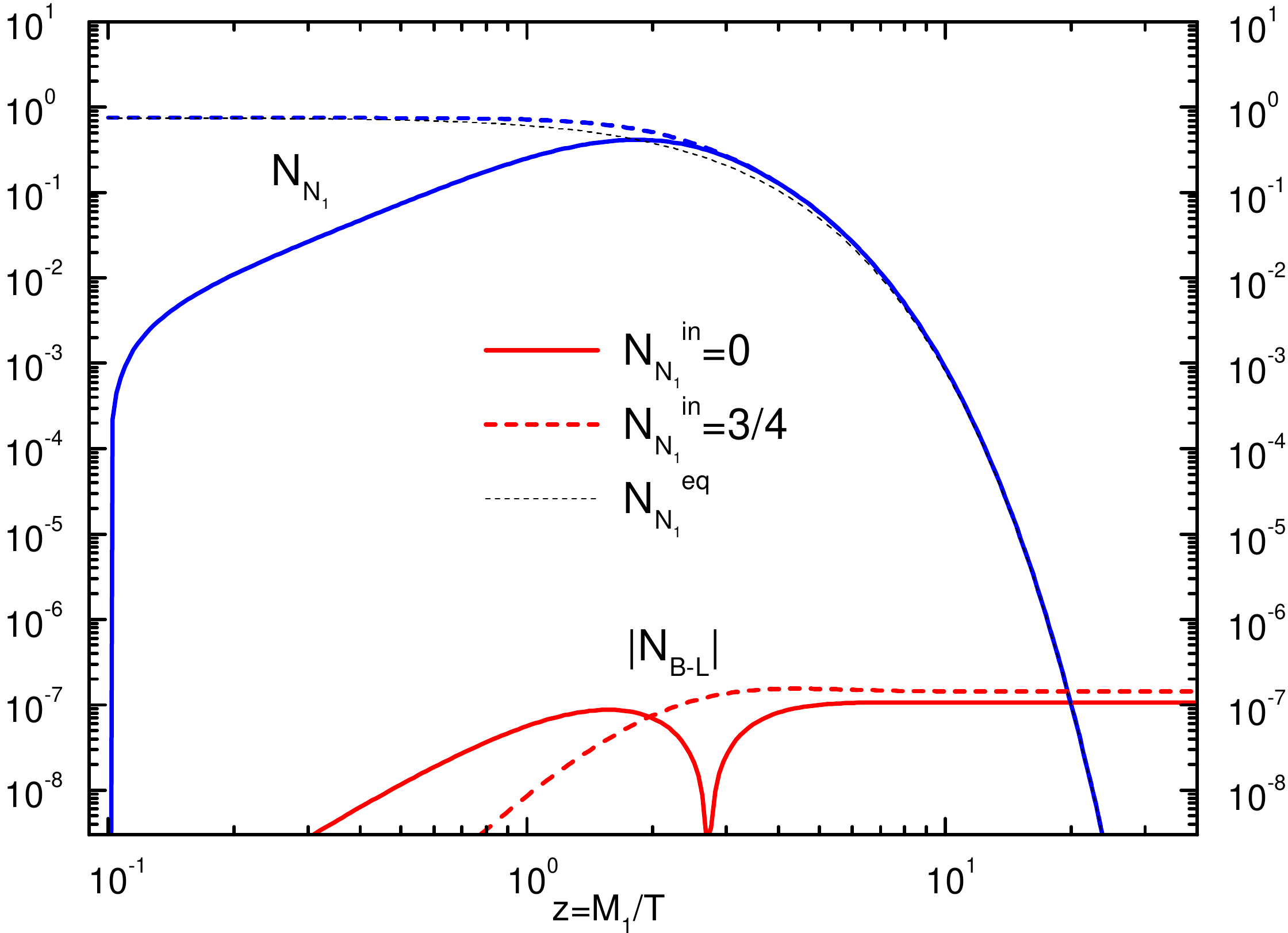}
	\caption{Evolution of heavy neutrino abundance ${\rm N_{N_1}}$
          and lepton asymmetry ${\rm N_{B-L}}$ for typical leptogenesis
          parameters: $M_1 = 10^{10}\ {\rm GeV}$, $\widetilde{m}_1 =
          8\pi\Gamma_1  (v_{\rm EW})/M_1)^2\ \mathrm{eV}$, $\epsilon_1 =
          10^{-6}$; the inverse temperature $z=M_1/T$ is the time
          variable. The dashed (full) lines correspond to thermal
          (vacuum) initial conditions for the heavy neutrino
          abundance; the dotted line represents the equilibrium abundance.
          From Ref.~\cite{Buchmuller:2002rq}.}
	\label{fig:example}
\end{center}
\end{figure}
Note that the $C\!P$ asymmetry $\epsilon_1$ results from a quantum
interference. On the contrary, washout terms, which are
neglected in Eqs.~(\ref{BE1}) and (\ref{BE2}), are tree level proesses.

Solutions of the Boltzmann equations are shown in Fig.~10 for
different initial $N$-distributions: thermal abundance and zero
abundance, respectively. It is important that the final 
$B$$-$$L$ asymmetry is essentially independent of the initial
conditions. This holds for sufficiently large values of the effective
light neutrino mass, $\widetilde{m}_1 \gtrsim 10^{-3}~{\rm eV}$. In
the case of hierarchical GUT scale neutrinos, 
$M_3 \sim \Lambda_{\rm GUT} \sim 10^{15}~{\rm GeV} \gg M_1 \sim
10^{10}~{\rm GeV}$, the required reheating temperature 
is $T_R \sim M_1 \sim 10^{10}~{\rm GeV}$, eight orders of
magnitude larger than the temperature of electroweak baryogenesis.

During the past years detailed studies have been caried out on bounds for
neutrino masses and mixings from leptogenesis. It is then important to
go beyond the ``one-flavour-approximation''. Furthermore, important results have
been obtained for specific lepton flavour models, in particular
in the context of GUT models, for different realizations of the seesaw
mechanism, and on the connection with $C\!P$
violation in low energy processes \cite{flavour}.

\section{Cosmological $B$$-$$L$ Breaking}

So far we have seen that the smallness of the light neutrino masses can
be explained by the seesaw mechanism, i.e. their mixing with
heavy Majorana neutrinos, and that $C\!P$-violating decays and
scatterings of these heavy neutrinos naturally yield the observed
baryon asymmetry. The heavy Majorana masses break $B$$-$$L$, and on
theoretical grounds one expects that they result from the spontaneous
breaking of a local symmetry. Furthermore, stabilizing the hierarchy
between the electroweak scale and the heavy neutrino masses suggests supersymmetry. 
One thus arrives at a supersymetric extension of the Standard Model
with right-handed neutrinos and local $B$$-$$L$ symmetry. It is
remarkable that this simple framework contains all the ingredients
which are needed to account also for dark matter and inflation, in
addition to the matter-antimatter asymmetry. In the following we shall
describe this scenario, closely following Ref.~\cite{Buchmuller:2012wn}.

As we have seen in the previous section, thermal leptogenesis requires
a rather large reheating temperature, $T_L \sim 10^{10}~\mathrm{GeV}$. In   
supersymmetric theories this causes a potential problem because of
gravitino production from the thermal bath
\cite{Khlopov:1984pf,Ellis:1984eq}, 
which yields the abundance \cite{Bolz:2000fu,Pradler:2006hh},
\begin{equation}
\Omega_{\tilde{G}} h^2 = C
\left(\frac{T_{R}}{10^{10}\,\textrm{GeV}}\right)
\left(\frac{100\,\textrm{GeV}}{m_{\tilde{G}}}\right)
\left(\frac{m_{\tilde{g}}}{1\,\textrm{TeV}}\right)^2 \ , 
\end{equation}
where $C \sim 0.5$,  and $T_{R}$ is the reheating temperature. For
unstable gravitinos, one has to worry about consistency with
primordial nucleosynthesis (BBN) whereas stable gravitinos may overclose the universe.
As a possible way out, nonthermal production of heavy neutrinos has
been suggested 
\cite{Lazarides:1991wu,Murayama:1992ua,Asaka:1999yd,Antusch:2010mv},
which allows to decrease the reheating temperature and therefore the
gravitino production. On the other hand, it is remarkable that for
typical gravitino and gluino masses in gravity mediated supersymmetry
breaking, a reheating temperature $T_{R} \sim 10^{10}\,\mathrm{GeV}$
yields the right order of magnitude for the dark matter abundance if
the gravitino is the LSP. But why should the reheating temperature be
as large as the temperature favoured by leptogenesis, i.e., $T_{R} \sim T_L$?

It this context it is interesting to note that for typical neutrino mass parameters in
leptogenesis, $\widetilde{m}_1 \sim 0.01\,\mathrm{eV}$,
$M_1 \sim 10^{10}\,\mathrm{GeV}$, the heavy neutrino decay width takes
the value
\begin{equation}
\Gamma_{N_1}^0 = 
\frac{\tilde{m}_1}{8 \pi} \left(\frac{M_1}{v_\textrm{\tiny EW}}\right)^2
\sim 10^3\ \textrm{GeV} \ . 
\end{equation}
If the early universe in its evolution would reach a state where the
energy density is dominated by nonrelativistic heavy neutrinos, their subsequent
decays to lepton-Higgs pairs would then yield a relativistic plasma with temperature 
\begin{equation}
T_{\rm R} \sim 0.2 \cdot \sqrt{\Gamma_{N_1}^0 M_P} \sim 10^{10}~\textrm{GeV} \ ,
\end{equation}
which is indeed the temperature wanted for gravitino dark matter! Is
this an intriguing hint or just a misleading coincidence?

\subsection{$B$$-$$L$ Breaking and False Vacuum Decay} 

We shall now demonstrate that an intermediate stage of heavy neutrino
dominance indeed
occurs in the course of the cosmological evolution if the initial
inflationary phase is driven by the false vacuum energy of unbroken
$B$-$L$ symmetry \cite{Buchmuller:2010yy,Buchmuller:2012wn}. 

Consider the supersymmetric standard model with right-handed
neutrinos, described by the superpotential (in $S\!U(5)$ notation:
 ${\bf 10} = (q,u^c,e^c)$, ${\bf 5^*} = (d^c,l)$),
\begin{eqnarray}\label{sbml}
W_M = h_{ij}^u {\bf 10}_i{\bf 10}_j H_u 
          +  h_{ij}^d {\bf 5}^*_i{\bf 10}_j H_d  \nonumber \\ 
          + h_{ij}^{\nu} {\bf 5}^*_i n^c_j H_u +  h_i^n n^c_i n^c_i S_1 \ ,
\end{eqnarray}
supplemented by a term which enforces $B$-$L$ breaking,
\begin{equation}\label{hinf}
W_{B-L} = \frac{\sqrt{\lambda}}{2} \Phi \left(v_{B-L}^2 - 2 S_1 S_2\right)\ .
\end{equation}
The Higgs fields $H_{u,d}$ and $S_{1,2}$ break electroweak symmetry
and $B$-$L$ symmetry, respectively, with $\langle H_{u,d}\rangle =
v_{u,d}$ and $\langle S_{1,2} \rangle = v_{B-L}/\sqrt{2}$.
It is well known that the superpotential $W_{B-L}$ can successfully
describe inflation with $\Phi$ as inflaton field, which is referred to
as F-term hybrid inflation\cite{Copeland:1994vg,Dvali:1994ms}. 

The Yukawa couplings in the superpotential $W_M$ are largely
determined by low energy physics of quarks, charged leptons and
neutrinos. It is useful to take this into account by means of a
specific flavour model. In the following we choose a model with $U(1)$
flavour symmetry of Froggatt-Nielsen (FN) type. Up to factors
$\mathcal{O}(1)$, the Yukawa couplings are given by 
\bea
h_{ij} \sim \eta^{Q_i + Q_j}\ , \quad \sqrt{\lambda} \sim
\eta^{Q_\Phi}\ , 
\eea
with $\eta \simeq 1/\sqrt{300}$.
The FN charges $Q_i$ are chosen following
Ref.~\cite{Buchmuller:1998zf} and listed in Table~2.
\begin{table}[t]
\begin{center}
\begin{tabular}{c|cccccccccccc}\hline 
$\psi_i$ & ${\bf 10}_3$ & ${\bf 10}_2$ & ${\bf 10}_1$ & ${\bf 5}^*_{3,2}$ & ${\bf 5}^*_1$ &
$n^c_{3,2}$ & $n^c_1$ & $H_{u,d}$ & $S_{1,2}$ & $\Phi$  \\ \hline
$Q_i$ & 0 & 1 & 2 & $a$ & $a+1$ & $d-1$ & $d$ & 0 & 0 & $2(d-1)$
\\ \hline
\end{tabular}
\end{center}
\caption{Assignement of FN charges of $U(1)$ flavour symmetry.} 
\end{table}

For simplicity, we will restrict our analysis to the case of a
hierarchical heavy (s)neutrino mass spectrum, $M_1 \ll M_{2}, M_3$,
where $M = h^n \, v_{B-L}$. Furthermore, we assume the heavier
(s)neutrino masses to be of the same order of magnitude as the common mass $m_S$ of the particles in the symmetry breaking sector, for definiteness we set $M_{2} = M_3 =  m_S$. 
Taking the $B$$-$$L$ gauge coupling to be $g^2 = g_{GUT}^2 \simeq \pi/6$, the model is, up to ${\cal O}(1)$ factors, determined by the $U(1)_{\text{FN}}$ charges $a$ and $d$. The $B$$-$$L$ breaking scale $v_{B-L}$, the mass of the lightest of the heavy (s)neutrinos $M_1$, and the effective light neutrino mass parameter $\widetilde{m}_1$ are related to these by
\begin{align}
\label{eq_v0}
 v_{B-L} &\sim \eta^{2a} \frac{v_{\text{EW}}^2}{\overline{m}_{\nu}} \
 , \quad M_1 \sim \eta^{2d} v_{B-L}\ , \\
\widetilde{m}_1 &= \frac{(m_D^{\dagger} m_D)_{11} }{M_1} \sim
\eta^{2a} \frac{v^2_{\text{EW}}}{v_{B-L}}\ ,
\end{align}
where $\overline{m}_{\nu} = \sqrt{m_2 m_3}$, the geometric mean of the two
light neutrino mass eigenvalues $m_2$ and $m_3$, characterizes the
light neutrino mass scale that, with the charge assignments above,
can be fixed to $3 \times 10^{-2}$~eV. Here, the seesaw formula 
$m_{\nu} = -m_D M^{-1} m_D^T$ has been exploited, with $m_D = h^{\nu}
v_{\text{EW}}$. Note, that $\widetilde{m}_1$ is bounded from below by
the lightest neutrino mass $m_1$ \cite{Fujii:2002jw}. Instead of the
FN $U(1)$ flavour charges the physical quantities $v_{B-L}$, $M_1$ and
$\widetilde{m}_1$ can be used as parameters of the model. In the
discussion of dark matter the gravitino ($m_{\tilde G}$) and gluino
($m_{\tilde g}$) masses enter as additional parameters.

Before the spontaneous breaking of $B$$-$$L$, supersymmetry is broken
by the vacuum energy density $\rho_0 = \frac{1}{4} \lambda v_{B-L}^4$,
which drives inflation. During this time, the dynamics of the system
is governed by the slowly rolling scalar component $\phi$ of the
inflaton multiplet~$\Phi$. The scalar components of the Higgs
superfields $S_{1,2}$ are stabilized at zero. As the field value of
the inflaton decreases, so do the effective masses in the Higgs
sector, until a tachyonic direction develops in the effective scalar
potential, which triggers a rapid transition to a phase with
spontaneously broken $B$$-$$L$ aymmetry (see Fig.~11). 

The phase transition is best treated in unitary gauge where the physical degrees of freedom are manifest. Performing a super-gauge transformation relates the Higgs superfields $S_{1,2}$ and the vector superfield $V$ to the respective fields $S'$ and $Z$ in unitary gauge,
\begin{figure}[t]
\begin{center}
\includegraphics[width=8.5cm,height=7cm]{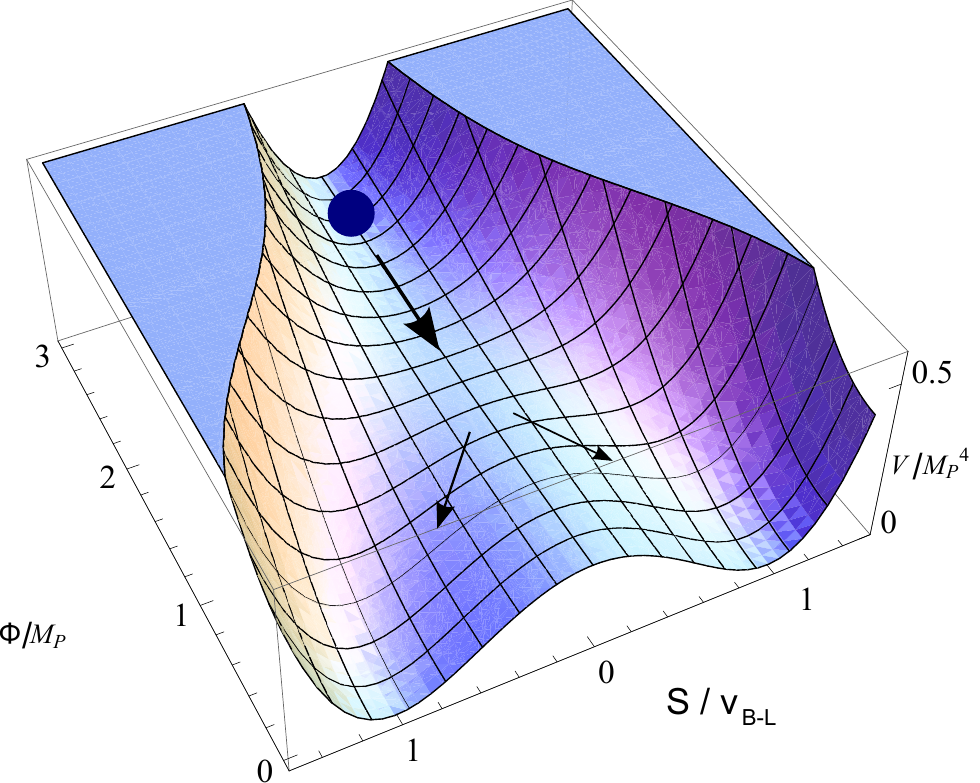}
%\parbox{17cm}{\epsfig{densities2.eps,width=16cm,height=9cm}}
\end{center}
\caption{Hybrid inflation: The time evolution of the inflaton field
  $\Phi$ leads to a tachyonic mass of the waterfall field $S$, which
  triggers a rapid transition to a phase with spontaneously broken
  $B$$-$$L$ symmetry.}
\end{figure}
\begin{equation}
\label{eq_S12}
 S_{1,2} = \frac{1}{\sqrt{2}} \, S' \, \exp (\pm i T) \,, \qquad V = Z + \frac{i}{2 g  q_S} (T - T^*) \,.
\end{equation}
The supermultiplet $S'$ contains two real scalar degrees of freedom,
$s' = \frac{1}{\sqrt 2}(\sigma' + i \tau)$, where $\tau$ remains
massive throughout the phase transition and $\sigma'$ is the symmetry
breaking Higgs field. It acquires a vacuum expectation value
proportional to $v(t) = \frac{1}{\sqrt{2}}\langle \sigma'^2(t,
\vec{x})\rangle_{\vec{x}}^{1/2}$ which approaches $v_{B-L}$ at large
times. In the Lagrangian, symmetry breaking is described by the replacement $\sigma' \rightarrow \sqrt{2} v(t) + \sigma$, where $\sigma$ denotes the fluctuations around the homogeneous Higgs background. 
The fermionic component $\tilde{s}$ of the supermultiplet $S'$ pairs
up with the fermionic component $\tilde{\phi}$ of the inflaton
supermultiplet $\Phi$ to form a Dirac fermion $\psi$, the higgsino,
which becomes massive during the phase transition. Due to
supersymmetry, the corresponding scalar fields ($\sigma$, $\tau$ and
inflaton $\phi$) have the same mass as the higgsino in the supersymmetric true vacuum.
Likewise, the gauge supermultiplet $Z$ (gauge boson $A$, real scalar
$C$, Dirac gaugino $\tilde A$) and the (s)neutrinos $N_i$
($\tilde{N_i}$) acquire masses. 

At the end of the phase transition, supersymmetry is restored.
An explicit calculation of the Lagrangian describing this phase
transition yields the time-dependent mass eigenvalues:
\begin{equation}
\label{eq_masses}
\begin{split}
&m^2_{\sigma} = \frac{1}{2} \lambda(3 v^2(t) - v_{B-L}^2) \,, \qquad
m^2_{\tau} = \frac{1}{2} \lambda (v_{B-L}^2 + v^2(t)) \ , \\
&m^2_{\phi} = \lambda v^2(t) \,, \qquad m^2_{\psi} = \lambda v^2(t) \,, \\ 
&m^2_Z = 8 g^2 v^2(t) \ , \\
&M^2_i = (h_i^n)^2 v^2(t) \ ,
\end{split}
\end{equation}
where we corrections due to thermal effects and
supersymmetry breaking have been ignored.

The symmetry breaking proceeds very rapidly, and therefore it is often referred to as a `waterfall' transition. It is accompanied by the production of local topological defects in the form of cosmic strings as well as the nonadiabatic production of particles
coupled to the Higgs field, a process commonly known as tachyonic
preheating~\cite{Felder:2000hj}. 

The cosmic strings produced during the phase transition have an energy
per unit length \cite{Hindmarsh:2011qj},
\begin{equation}
\label{eq_mu}
\mu = 2 \pi B(\beta) v_{B-L}^2 \,,
\end{equation}
with $\beta = \lambda / (8 \, g^2)$ and $B(\beta) = 2.4 \left[ \ln(2/\beta) \right]^{-1}$ for $\beta < 10^{-2}$. According to~Ref.~\cite{Copeland:2002ku}, the characteristic length separating two strings formed during tachyonic preheating is
\begin{equation}
\label{eq_xi}
\xi = (- \lambda v_{B-L} \dot{\varphi}_c)^{-1/3} \,.
\end{equation}
Here $\dot{\varphi}_c$ is the velocity of the radial component of the
inflaton field, $\phi = \varphi / \sqrt{2} e^{i \theta}$, at the onset
of the phase transition, which can be determined from the scalar
potential using the equation of motion for $\varphi$.
In the region of parameter space we are interested in, the slope of
the scalar potential is determined by the one-loop quantum corrections
(cf., e.g., Ref.~\cite{Nakayama:2010xf}). 
 With this, one obtains for the energy density stored in strings just after the end of the phase transition
\begin{equation}
\label{eq_rhos}
\rho_{\text{string}} = \frac{\mu}{\xi^2} \ . 
\end{equation}
From Eqs.~\eqref{eq_mu}, \eqref{eq_xi} and the one-loop potential, one
finds that the fraction of energy stored in cosmic strings directly
after the phase transition increases strongly with the coupling
$\lambda$. This is due to the higher energy density per cosmic string
as well as the shorter average distance between two strings. For
instance, for $v_{B-L} = 5 \times 10^{15}$~GeV and $\lambda =
10^{-2}$, one has $(H \xi)^{-1} \simeq 400$ and $\rho_{\text{string}}/ \rho_0 \simeq 60\,\%$. For $\lambda = 10^{-5}$, this is reduced to $(H \xi)^{-1} \simeq 40$ and $\rho_{\text{string}}/ \rho_0 \simeq 0.2\,\%$.

These relic cosmic strings can in principal be observed today, e.g.\ via string induced gravitational lensing effects in the CMB.
The nonobservation of these effects implies an upper bound on the energy per unit length~\cite{Battye:2010xz,Urrestilla:2011gr,Dvorkin:2011aj},
\begin{equation}
G \mu \lesssim 5 \times 10^{-7} \,,
\end{equation}
where $G =  \Mp^{-2}$ is Newton's constant with $\Mp = 1.22 \times
10^{19}$~GeV denoting the Planck mass. Inserting this into
Eq.~\eqref{eq_mu} puts an upper bound on $v_{B-L}$.  In
Ref.~\cite{Nakayama:2010xf}, also the  bounds inferred from the
spectrum of fluctuations in the CMB~\cite{Komatsu:2010fb} have been
taken into account, yielding the viable parameter range
\begin{equation}
\begin{split}
&3 \times 10^{15}~\text{GeV} \; \lesssim v_{B-L} \lesssim  \; 7 \times 10^{15}~\text{GeV} \,, \\
&10^{-4} \; \lesssim  \sqrt{\lambda}  \lesssim \; 10^{-1} \,. 
\end{split}
\end{equation}
This significantly constrains the model parameters. With the
scale of $B$$-$$L$ breaking basically fixed, one finds for the FN flavour charges
$a = 0$ and $1.4 \lesssim d \lesssim 2.6$, corresponding to the range
\begin{equation}
\begin{split}
& 10^9~\text{GeV} \leq M_1 \leq 3 \times 10^{12}~\text{GeV} \ ,  \\
& 10^{-5}~\text{eV} \leq \widetilde{m}_1 \leq 1~\text{eV} \ .
\end{split}
\label{eq_parameter_space}
\end{equation} 

During tachyonic preheating, quantum fluctuations of the Higgs field $\sigma'_k$ with wave number $|\vec{k}| < |m_{\sigma}|$ grow exponentially, while its average value remains zero. The strong population of the long wavelength Higgs modes leads to a large abundance of nonrelativistic Higgs bosons. Other particles coupled to the Higgs field are nonperturbatively produced due to the rapid change of their effective masses~\cite{GarciaBellido:2001cb}.

The mode equations for the gauge, Higgs, inflaton, and neutrino supermultiplets are governed by the time-dependent masses proportional to $v(t)$ listed in Eq.~\eqref{eq_masses}. This leads to particle production \cite{GarciaBellido:2001cb}, with number densities for bosons and fermions after tachyonic preheating given by\footnote{Note that particle production can be significantly enhanced by quantum effects \cite{Berges:2010zv}, which require further investigations.}
\begin{align}
&n_B(\alpha) \simeq 1 \times 10^{-3} g_s m_S^3
f(\alpha, 1.3)/\alpha \ , \nonumber \\
&n_F(\alpha) \simeq 3.6 \times 10^{-4} g_s
m_S^3 f(\alpha, 0.8)/\alpha \ , \label{eq_partprod}
\end{align}
with $f(\alpha, \gamma) = \sqrt{\alpha^2 + \gamma^2} - \gamma$ and $\alpha = m_X/m_S$, where $m_X$ denotes the mass of the respective particle in the true vacuum; $g_s$ counts the spin degrees of freedom of the respective particle. Just as the Higgs bosons themselves, these particles are produced with very low momenta, i.e.\ nonrelativistically.

\begin{figure}
\begin{center}
\hspace*{-0.5cm}
\includegraphics[width=6.5cm,height=5cm]{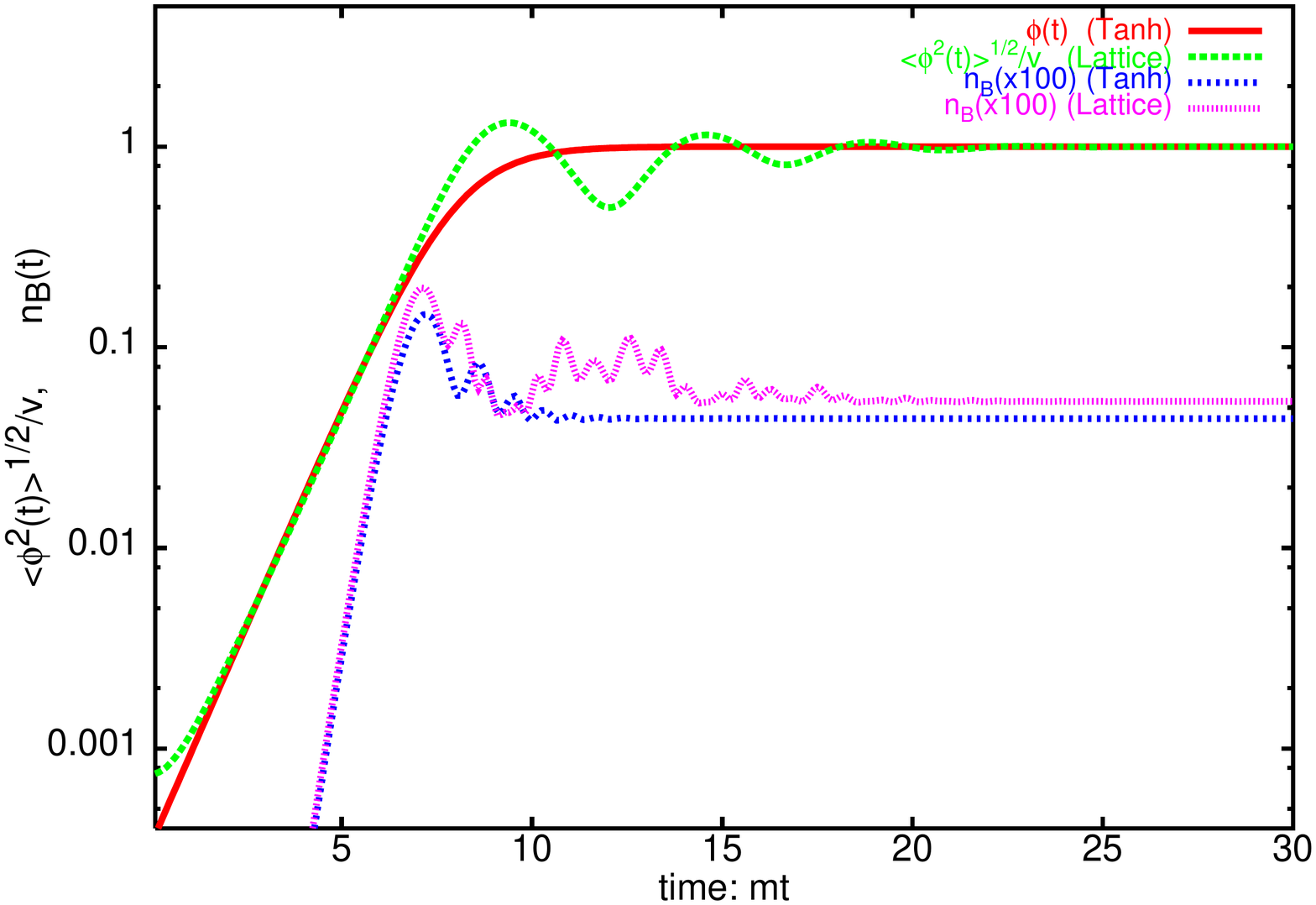}\hspace*{0.5cm}
\includegraphics[width=6.5cm,height=5cm]{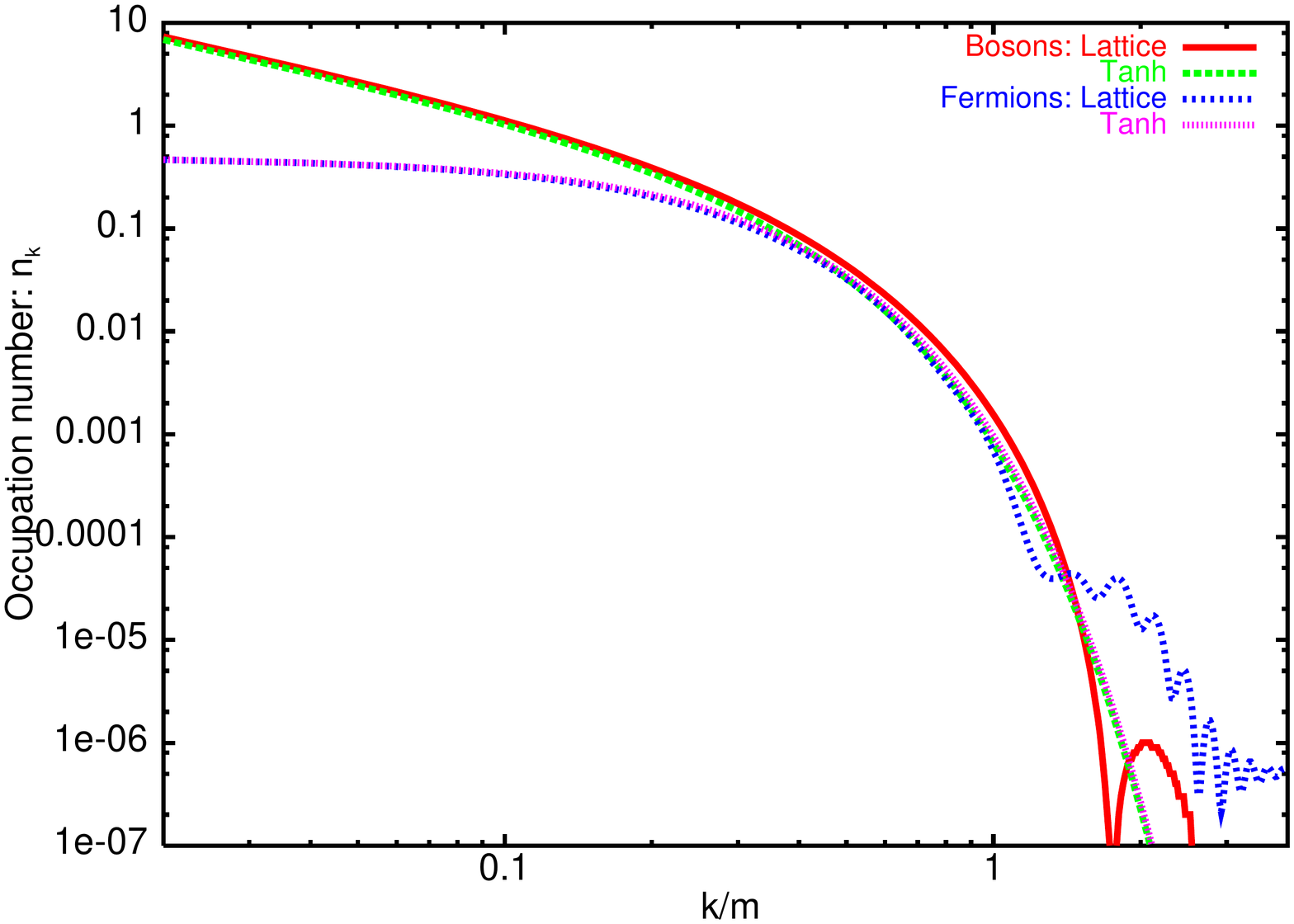}
%\parbox{17cm}{\epsfig{densities2.eps,width=16cm,height=9cm}}
\end{center}
\caption{(a) (left) Growth of the $B$$-$$L$ Higgs expectation value as
  function of time. (b) (right) Occupation numbers of bosons and
  fermions produced during tachyonic preheating as functions of
  momentum. From Ref.~\cite{GarciaBellido:2001cb}.}
\end{figure}

\subsection{The Reheating Process}

During tachyonic preheating, most of the vacuum energy is converted
into Higgs bosons ($\sigma$). At the same time, particles coupled to
the Higgs field, i.e.\ quanta of the gauge, Higgs, inflaton and
neutrino supermultiplets are produced, with the resulting abundances
given by Eq.~\eqref{eq_partprod}. Among these particles, the members
of the gauge supermultiplet have by far the shortest lifetime. Due to
their large couplings they decay basically instantaneously into
(s)neutrinos and MSSM particles. This sets the initial conditions for
the following phase of reheating, which can be described by means of Boltzmann equations.

Due to the choice of the hierarchical (s)neutrino mass spectrum, the decay of particles from the symmetry breaking sector into the two heavier (s)neutrino generations is kinematically forbidden. These particles can hence only decay into particles of the $N_1$ supermultiplet. These (s)neutrinos, just as the neutrinos produced through gauge particle decays and thermally produced (s)neutrinos, decay into MSSM particles, thereby generating the entropy of the thermal bath as well as a lepton asymmetry. Note that these different production mechanisms for the (s)neutrinos yield (s)neutrinos with different energies, which due to relativistic time-dilatation, decay at different rates. Finally, the thermal bath produces a thermal gravitino density, which turns out to be in the right ball-park to yield the observed dark matter abundance.

The network of Boltzmann equations for the time evolution of all
particles and superparticles is described in detail in
Ref.~\cite{Buchmuller:2012wn}. Their solution provides the initial
conditions of the hot early universe. In Fig.~13 the result is shown
for an illustrative choice of parameters,
\begin{align}
M_1 &= 5.4 \times 10^{10}\,\textrm{GeV}\ ,\quad
\widetilde{m}_1 = 4.0 \times 10^{-2}\,\textrm{eV}\ ,\nonumber\\
m_{\widetilde{G}} &= 100\,\textrm{GeV}\ ,\quad
m_{\tilde{g}} = 1\,\textrm{TeV}\ .
\end{align}
\begin{figure}
\begin{center}
\includegraphics[width=10cm]{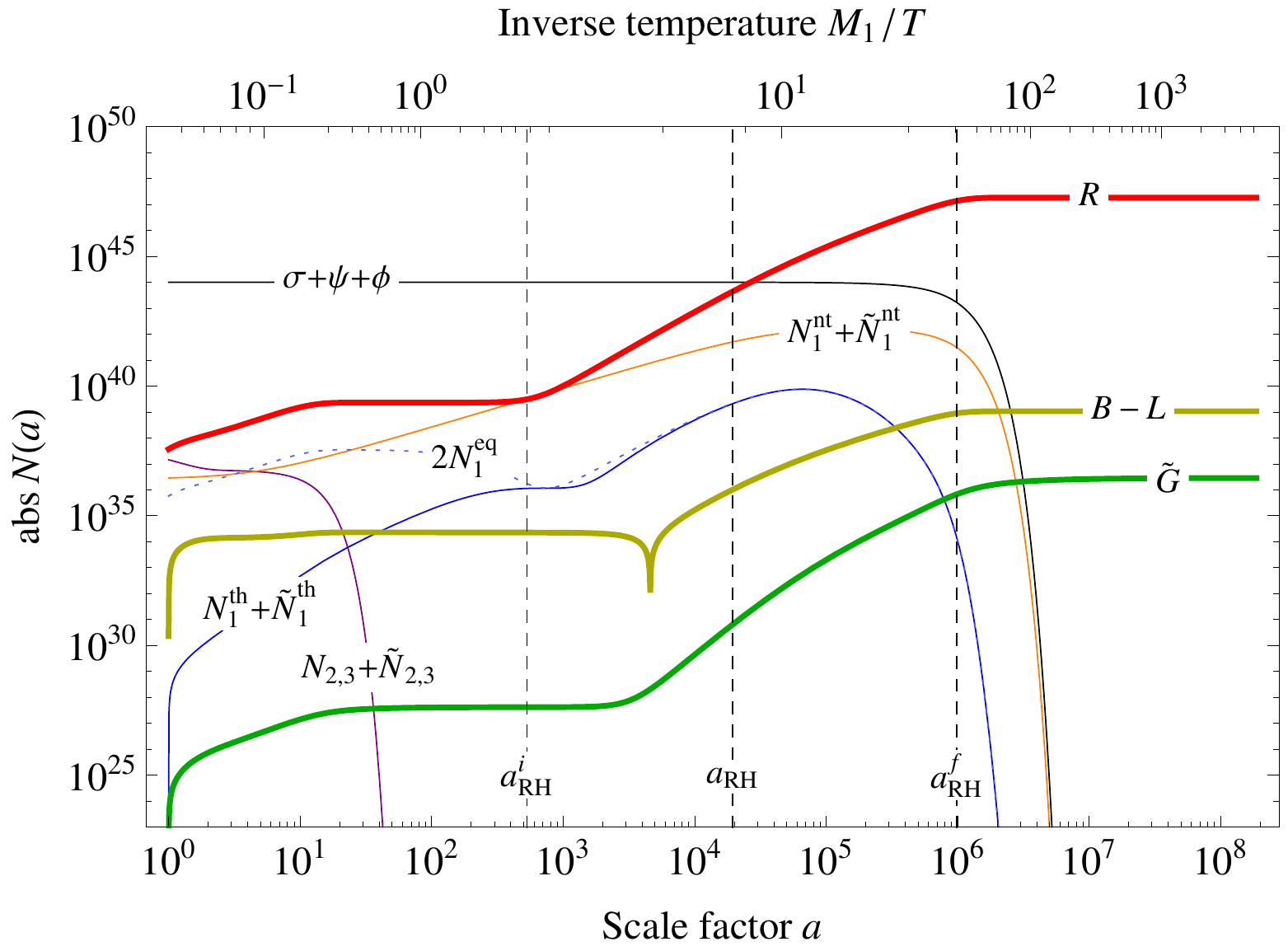}
\caption{Comoving number densities for particles from the symmetry breaking sector
(Higgs $\sigma$ + higgsinos $\psi$ + inflatons $\phi$), (non)thermally
produced (s)neutrinos of the first generation
($N_{1}^{\textrm{th}} + \tilde{N}_{1}^{\textrm{th}}$, $N_{1}^{\textrm{nt}} + \tilde{N}_{1}^{\textrm{nt}}$),
(s)neutrinos of the first generation in thermal equilibrium
($2 N_1^\textrm{eq}$, for comparison),
(s)neutrinos of the second and third generation
($N_{2,3} + \tilde{N}_{2,3}$), the MSSM radiation ($R$),
the lepton asymmetry ($B$$-$$L$), and gravitinos ($\widetilde{G}$)
as functions of the scale factor $a$.
The vertical lines labeled $a_{\textrm{RH}}^i$, $a_{\textrm{RH}}$ and $a_{\textrm{RH}}^f$
mark the beginning, the middle and the end of the reheating process.
From Ref.~\cite{Buchmuller:2012wn}.}
\label{fig:numengden}
\end{center}
\end{figure}
Due to the rapid decay of gauge particles, the total energy density has
already right after the end of tachyonic preheating a small
relativistic component. For a rather long period, during which the
scale factor grows by six orders of magnitude, it is dominated by the
nonrelativistiv gas of Higgs bosons. Their decay via heavy neutrinos
then generates the bulk of entropy, baryon asymmetry and gravitino abundance.
For the chosen parameters, dark matter is made of gravitinos. Their
abundance and the baryon-to-photon ratio are consistent with observation,
\bea
\eta_B \simeq 3.7 \times 10^{-9}  \ , \quad
\Omega_{\widetilde{G}} h^2 \simeq 0.11\ .
\eea
Note that the given baryon-to-photon ratio corresponds to maximal
$C\!P$ asymmetry, which can be reduced by an appropriate choice of phases in the
neutrino mass matrices. 
\begin{figure}
\begin{center}
\includegraphics[width=9cm]{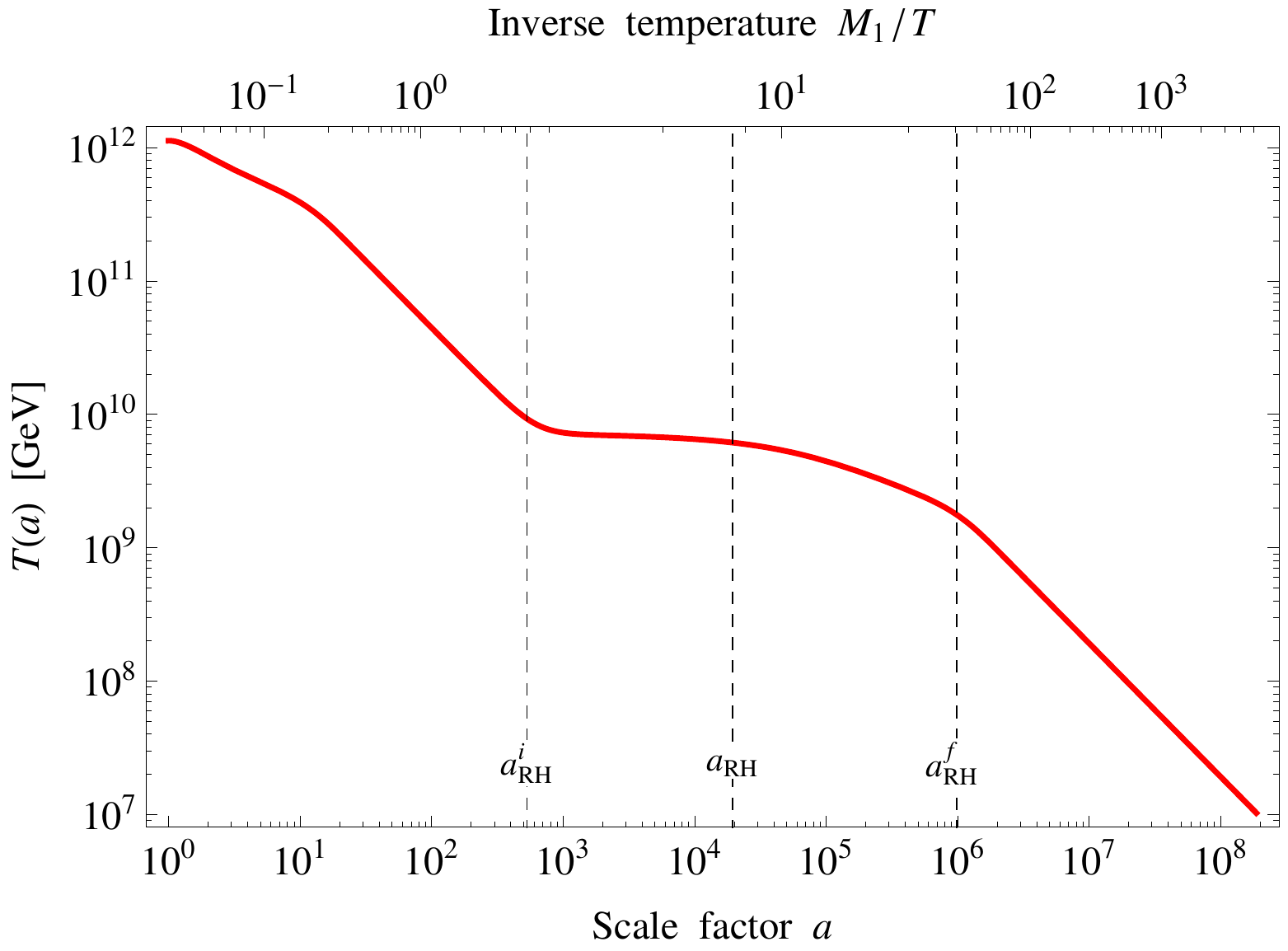}
\end{center}
\caption{Temperature $T$ of the thermal bath as function of the scale
  factor $a$. From Ref.~\cite{Buchmuller:2012wn}.}
\end{figure}

A key feature of the described reheating process is the emergence of
an approximate temperature plateau between $a_{\textrm{RH}}^i$ and
$a_{\textrm{RH}}^f$. The corresponding temperature 
$T_R \equiv T(a_{\rm  RH})$ is determined by neutrino masses,
\begin{align}
T_{R} \simeq 1.3 \times 10^{10}\,\textrm{GeV}
\left(\frac{\widetilde{m}_1}{0.04\,\textrm{eV}}\right)^{1/4}
\left(\frac{M_1}{10^{11}\,\textrm{GeV}}\right)^{5/4} \ .
\label{eq:TRHfit}
\end{align}
The temperature plateau occurs as result of a competition between universe
expansion and entropy production in $N_1$ decays. During this period
most of the baryon asymmetry and dark matter are produced.

\subsection{Leptogenesis and Dark Matter} 

In the described reheating process, baryogenesis is a mixture of
nonthermal and thermal leptogenesis, which considerably extends the
viable range in the  $M_1-\widetilde{m}_1$ plane compared to thermal
leptogenesis. Gravitino production is dominated by thermal
processes. A systematic parameter scan shows that gravitino dark
matter is possible in the mass range 
$10\ \textrm{GeV} \lesssim m_{\widetilde{G}} \lesssim 700\ \textrm{GeV}$
(assuming  $m_{\widetilde{g}} \simeq 1\ \textrm{TeV}$) (cf.~Fig.~15).
Gravitino dark matter further constrains the heavy neutrino mass
\begin{figure}
\begin{center}
\includegraphics[width=11cm,height=8cm]{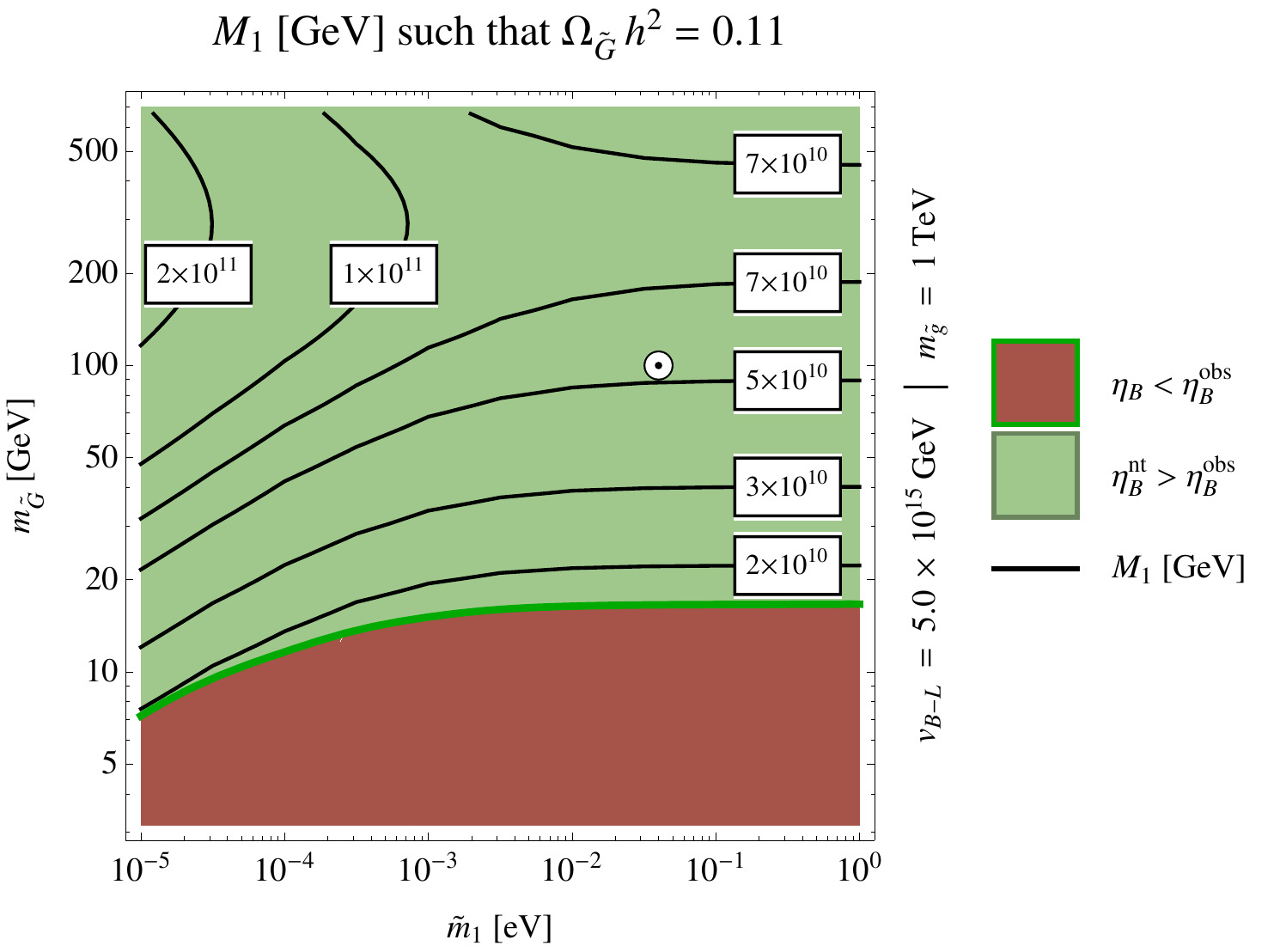}
\end{center}
\caption{Contour plots of the heavy neutrino mass $M_1$ as function
of the effective neutrino mass~$\widetilde{m}_1$ and the gravitino
mass $m_{\widetilde{G}}$
such that the relic density of dark matter is accounted for by gravitinos. 
In the red region the lepton asymmetry generated by leptogenesis is smaller than the
observed one, providing a lower bound on the gravitino mass, depending
on $\widetilde{m}_1$.  From Ref.~\cite{Buchmuller:2012wn}.
\label{fig:mGbounds}
}\end{figure}
to the range
$2\times 10^{10}\ \textrm{GeV} \lesssim M_1 \lesssim 2\times 10^{11}\ \textrm{GeV}$,
which is more stringent than the constraint from inflation.

Recent results on the Higgs boson mass from the LHC motivate a
superparticle mass spectrum with a very heavy gravitino
\cite{Ibe:2011aa,Jeong:2011sg},
\begin{equation}\label{masshierarchy}
m_{\mathrm{LSP}}  \ll 
m_{\mathrm{squark},\mathrm{slepton}} 
\ll m_{\widetilde{G}} \ .
\end{equation}
Due to this hierarchy the LSP is typically a `pure' gaugino or higgsino.
A pure neutral wino or higgsino is almost mass degenerate with a chargino
belonging to the same $\mathrm{SU(2)}$ multiplet. Hence, the current lower
bound on chargino masses also applies to the LSP, 
$m_{\mathrm{LSP}} \geq 94~\mathrm{GeV}$ \cite{Beringer:1900zz}.
It is well known that a gravitino heavier than $10\ \mathrm{TeV}$ can
be consistent with primordial nucleosynthesis as well as leptogenesis
\cite{Gherghetta:1999sw,Ibe:2004tg}.

The thermal abundance of a pure wino ($\widetilde w$) or higgsino ($\widetilde h$) LSP becomes significant for 
masses above $1~\mathrm{TeV}$, where it is well approximated
by \cite{ArkaniHamed:2006mb,Hisano:2006nn,Cirelli:2007xd}
\begin{equation}\label{dmth}
\Omega^{\mathrm{th}}_{\widetilde{w},\widetilde{h}} h^2 
= c_{\widetilde{w},\widetilde{h}} 
\left(\frac{m_{\widetilde{w},\widetilde{h}}}{1~\mathrm{TeV}}\right)^2 \ , 
\quad c_{\widetilde{w}} = 0.014\ , \quad c_{\widetilde{h}} = 0.10\ ,
\end{equation}
for wino and higgsino, respectively.

Consider now gravitino masses in the range from 
$10~\mathrm{TeV}$ to $10^3~\mathrm{TeV}$. 
The gravitino lifetime is given by
\begin{equation}\label{Glife}
\tau_{\widetilde{G}} = 
\left(\frac{1}{32\pi}\left(n_v + \frac{n_m}{12}\right)
\frac{m_{\widetilde{G}}^3}{M_{\mathrm{P}}^2}\right)^{-1}
= 24 \left(\frac{10~\mathrm{TeV}}{m_{\widetilde{G}}}\right)^3 \mathrm{s}\ ,
\end{equation}
where $n_v = 12$ and $n_m = 49$ are the number of vector and chiral
matter multiplets in the MSSM, respectively.
The lifetime (\ref{Glife}) corresponds to the decay temperature
\begin{equation}
T_{\widetilde{G}} = \left(\frac{90~M_{\mathrm{P}}^2}{\pi^2 
g_*(T_{\widetilde{G}})\tau^2_{\widetilde{G}}}\right)^{1/4}
= 0.24 \left(\frac{10.75}{g_*(T_{\widetilde{G}})}\right)^{1/4}
\left(\frac{m_{\widetilde{G}}}{10~\mathrm{TeV}}\right)^{3/2} \mathrm{MeV}\,,
\end{equation}
with $g_*(T_{\tilde G})= 43/4$ counting the effective number of relativistic degrees of freedom.
For gravitino masses between $10~\mathrm{TeV}$ to $10^3~\mathrm{TeV}$
the decay temperature $T_{\widetilde{G}}$ varies between 
$0.2~\mathrm{MeV}$ and $200~\mathrm{MeV}$, i.e.\ roughly between the
temperatures of nucleosynthesis and the QCD phase transition. In this
temperature range the entropy increase due to gravitino decays and hence the corresponding dilution of the baryon asymmetry are negligible.

The decay of a heavy gravitino, $m_{\widetilde{G}} \gg m_{\mathrm{LSP}}$,
produces approximately one LSP. This yields the nonthermal
contribution to dark matter \cite{Buchmuller:2012bt}
\begin{equation}\label{dmG}
\Omega_{\mathrm{LSP}}^{\widetilde{G}} h^2  =  \frac{m_{\mathrm{LSP}}}{m_{\widetilde{G}}}
\Omega_{\widetilde{G}} h^2 
 \simeq 2.7\times 10^{-2} 
\left(\frac{m_{\mathrm{LSP}}}{100~\mathrm{GeV}}\right)
\left(\frac{T_{R}(M_1,\widetilde{m}_1)}{10^{10}~\mathrm{GeV}}
\right)\ ,
\end{equation}
where the reheating temperature is given by Eq.~(\ref{eq:TRHfit}).
For LSP masses below $1~\mathrm{TeV}$, which are most interesting for
the LHC as well as for direct searches, the total LSP abundance
\begin{equation}\label{dmtot}
\Omega_{\widetilde{w},\widetilde{h}} h^2 =
\Omega_{\widetilde{w},\widetilde{h}}^{\widetilde{G}} h^2 + 
\Omega_{\widetilde{w},\widetilde{h}}^{\mathrm{th}} h^2
\end{equation} 
is thus dominated by the contribution from gravitino decays.

The requirement of higgsino/wino dark matter, i.e.\
$\Omega_{\mathrm{LSP}} h^2 = \Omega_{\mathrm{DM}} h^2 \simeq 0.11$,
implies an upper bound on the reheating temperature, 
$T_{R} < 4.2\times 10^{10}~\mathrm{GeV}$. A lower bound
on $T_{R}$ is obtained from leptogenesis, depending on
$\widetilde{m}_1$ (cf.\ Fig.\ 16(a)).
\begin{figure}
\begin{center}
\hspace*{-0.5cm}
\includegraphics[width=6cm,height=5cm]{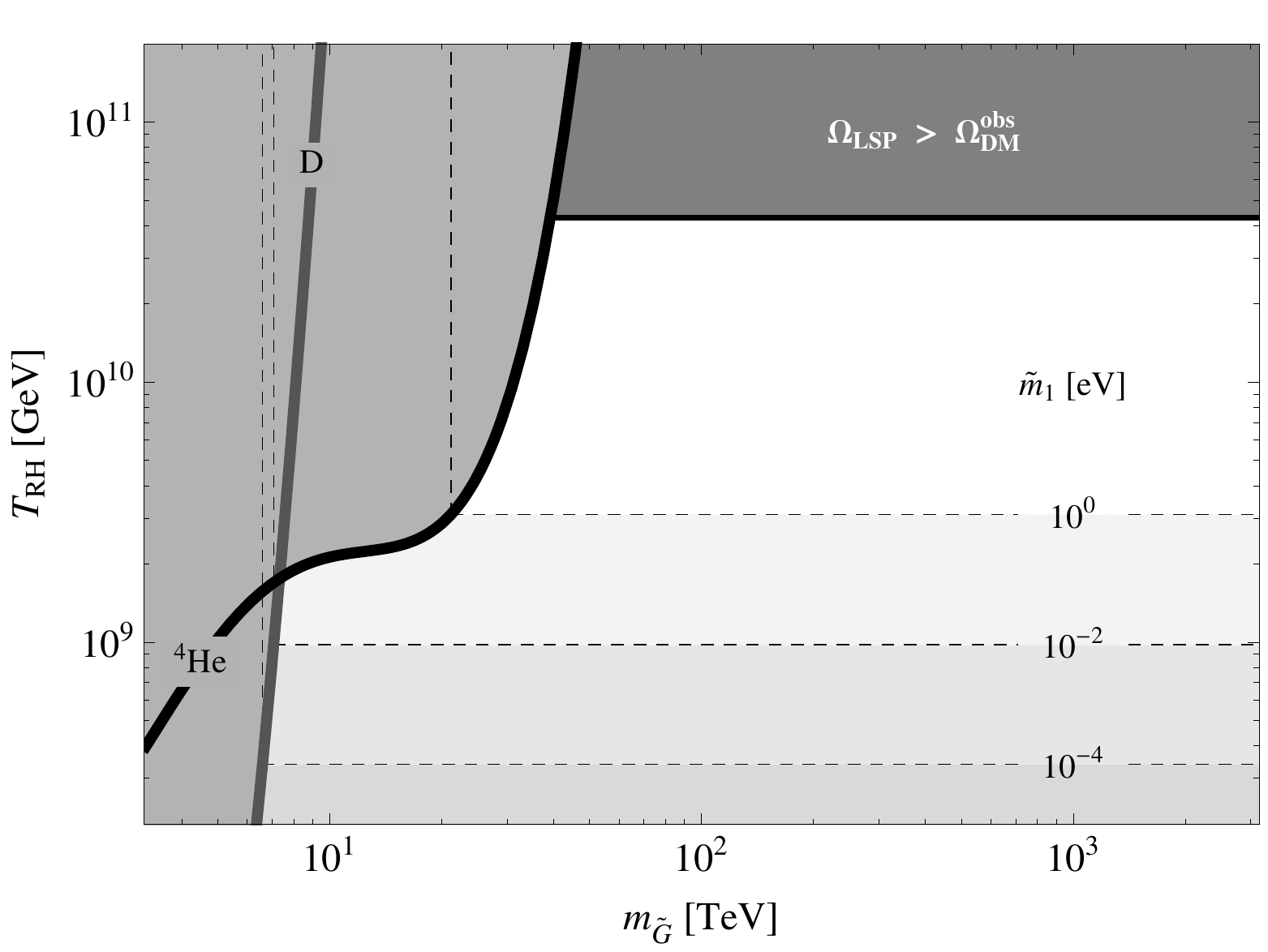}\hspace*{1.5cm}
\includegraphics[width=6cm,height=5cm]{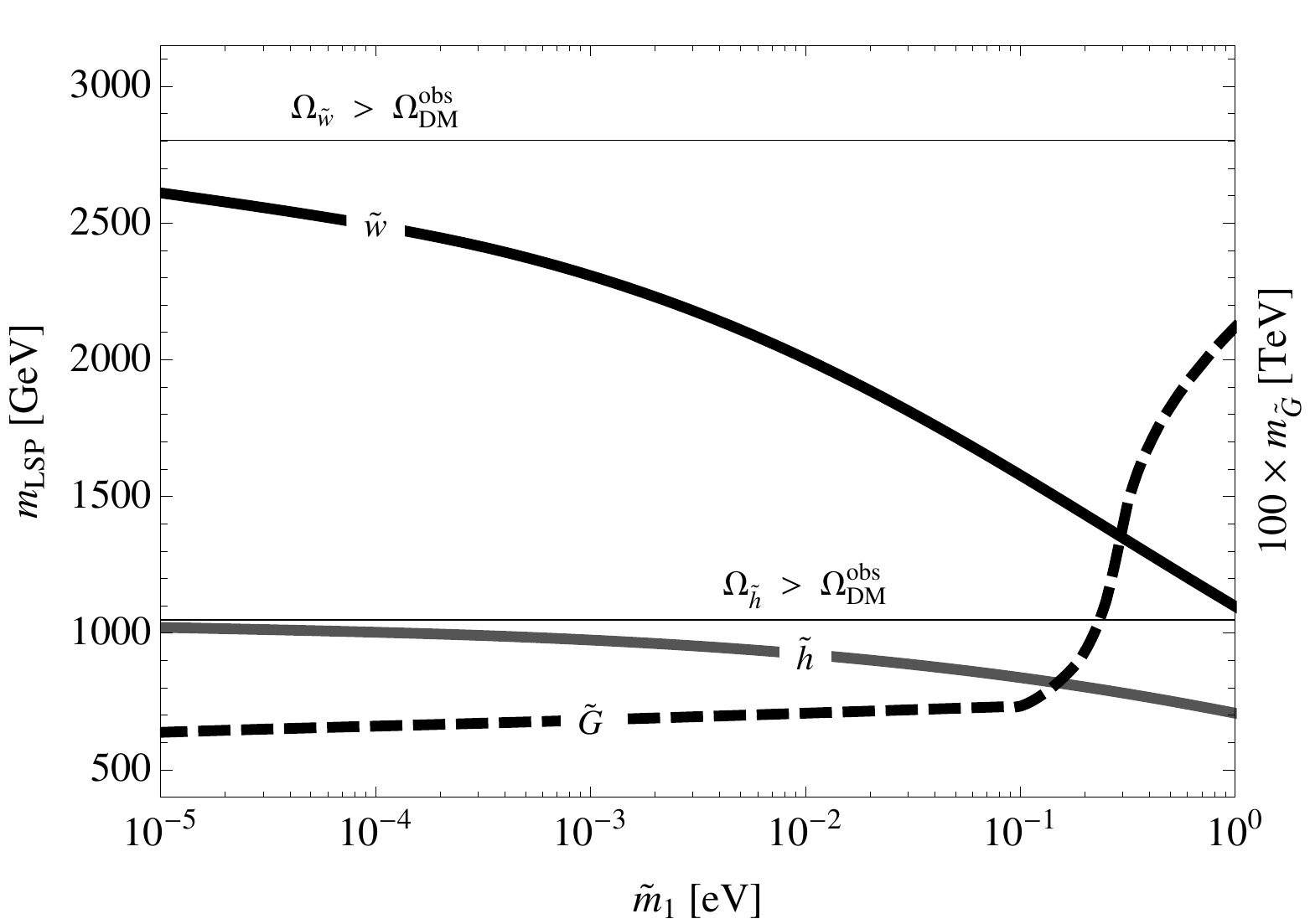}
\end{center}
\caption{(a) (left) Upper and lower bounds on the reheating
temperature as functions of the gravitino mass.
The horizontal dashed lines denote lower bounds imposed by
leptogenesis for different values of the effective neutrino mass
$\widetilde m_1$; the curves labelled $^4\text{He}$
and D denote upper bounds originating from the primordial helium-4 and deuterium abundances created during BBN \cite{Kawasaki:2008qe}.
The shaded region marked $\Omega_{\text{LSP}} >
\Omega^{\text{obs}}_{\text{DM}}$ is excluded by overproduction of dark
matter. 
(b) (right) Upper bounds on wino ($\widetilde{w}$) and higgsino
($\widetilde{h}$) LSP masses imposed by successful leptogenesis as well as absolute
lower bound on the gravitino mass according to BBN
as functions of the effective neutrino mass $\widetilde{m}_1$.
From Ref.~\cite{Buchmuller:2012bt}.}
\end{figure}
Higgsino/wino dark matter also implies an upper bound on the LSP mass,
depending on $\widetilde{m}_1$, and a lower bound on the gravitino
mass (cf.~Fig.~16(b)). For instance, for $m_1 = 0.05$~eV, one has  
$m_{\widetilde{h}} \lesssim 900$~GeV, $m_{\widetilde{G}} \gtrsim 10$~TeV. 

In summary, in the described scenario of cosmological $B$$-$$L$
breaking, the reheating temperature can vary in the range
$3\times 10^8\ \mathrm{GeV} \lesssim T_R \lesssim 5\times 10^{10}\
  \mathrm{GeV}$, depending on the nature of dark matter,
  i.e. gravitino or higgsino/wino.

\section{Conclusions and Outlook}

In the previous sections we have discussed several mechanisms for the
generation of matter and dark matter, which differ significantly with
respect to the theoretical framework, the predictive power and the
required reheating temperature in the early universe:  

\begin{itemize}
\item $T_R = \mathcal{O}(100~\mathrm{MeV})$: A reheating temperature
  just above the temperature where nucleosynthesis starts, is
  sufficient to generate baryon asymmetry and dark matter in moduli
  decay. The existence of such scalar fields is a generic feature of
  string compactifications. The model predicts nonthermal WIMP dark
  matter. The values of baryon asymmetry and dark matter abundance
  cannot be predicted since they depend on unknown moduli couplings.
\item $T_R = \mathcal{O}(100~\mathrm{GeV})$: Electroweak baryogenesis
  is a generic prediction of the Standard Model, which makes use of
  the electroweak phase transition and sphaleron processes in the
  high-temperature phase. However, due to the rather large 
  Higgs mass realized in nature, electroweak baryogenesis does not work in the
  Standard Model, not even in its supersymmetric extension. It remains
  a viable option in  case of a strongly interacting Higgs sector,
  which will be tested at the LHC.
\item $T_R = \mathcal{O}(100~\mathrm{GeV})$: The $\nu$MSM is indeed the
  most minimal extension of the Standard Model, which can account for
  both, baryogenesis and dark matter. However, a judicious choice of
  neutrino masses and mixings is required, which appears difficult to
  justify theoretically. The model can be verified or falsified by
  collider experiments and astrophysical observations in the near future. 
\item $T_R = \mathcal{O}(10^{10}~\mathrm{GeV})$: Thermal leptogensis in
  its simplest version explains the baryon asymmetry in terms of
  neutrino masses and mixings that are consistent with GUT
  model building. Dark matter must have another origin. Standard
  WIMP dark matter is incompatible with thermal leptogenesis. 
\end{itemize} 

Finally, we have described how spontaneous $B$$-$$L$ breaking together
with supersymmetry can account for baryon asymmetry, dark matter and
inflation. The reheating temperature $T_R$ can vary from
$\mathcal{O}(10^8 \mathrm{GeV})$ to $\mathcal{O}(10^{11} \mathrm{GeV})$. 
This simple picture is naturally consistent with neutrino physics and
GUT models. During the coming years we can hope to learn from
LHC data and astrophysical observations whether matter
and dark matter are remnants of the very early universe at temperatures
$\mathcal{O}(100~\mathrm{GeV})$, or whether temperatures several orders
of magnitude larger are needed, a possibility that is favoured by the idea of
grand unification.

\section*{Acknowledgments}
\noindent
The author thanks the organizers for their hospitality, Valerie Domcke and
Kai Schmitz for enjoyable collaboration on the topic of these
lectures, and Thomas Konstandin for comments on the manuscript.
This work has been supported by the German Science Foundation (DFG) within 
the Collaborative Research Center 676 ``Particles, Strings and the Early
Universe''.

\newpage

%% --- EDIT Bibriography ---

\end{document}